%% file: main.tex
\documentclass[10pt,journal,compsoc]{IEEEtran}
\usepackage{xcolor,soul,framed, caption} %,caption

\colorlet{shadecolor}{yellow}
\usepackage[pdftex]{graphicx}
\graphicspath{{../pdf/}{../jpeg/}}
\DeclareGraphicsExtensions{.pdf,.jpeg,.png}

\usepackage[cmex10]{amsmath}
\usepackage{array}
\usepackage{mdwmath}
\usepackage{mdwtab}
\usepackage{eqparbox}
\usepackage{url}
\usepackage{fontawesome}
\usepackage{booktabs}
\usepackage[percent]{overpic}

\usepackage{stfloats}
\usepackage{mathrsfs}
\usepackage{comment}
\usepackage{amsmath,amssymb,amsfonts,amsthm}
\usepackage{algorithmic}
\usepackage{textcomp}
\usepackage{xcolor}
\usepackage{subcaption}
\usepackage{listings}
\usepackage{multirow}

\usepackage{todonotes}
\usepackage[flushmargin]{footmisc}
\usepackage{wrapfig}
\usepackage{hyperref} % links
\usepackage[skip=10pt]{caption}
\usepackage{multicol}
\usepackage{hhline}
\usetikzlibrary{calc}
\usepackage{adjustbox}
\usepackage{pgfplots}

\newcounter{example}
\newenvironment{example}[1][]{\refstepcounter{example}\par\medskip\noindent
   \textbf{Example~\theexample. #1} \rmfamily}{\qed\medskip}

\definecolor{lightgray}{rgb}{.9,.9,.9}
\definecolor{darkgray}{rgb}{.4,.4,.4}
\definecolor{purple}{rgb}{0.65, 0.12, 0.82}

\def \repo{https://github.com/CSecLab/\toolname{}/}

\lstdefinelanguage{JavaScript}{
  basicstyle=\ttfamily\small,
  keywords={typeof, new, true, false, catch, function, return, null, catch, switch, var, if, in, while, do, else, case, break, const},
  keywordstyle=\color{blue}\bfseries,
  ndkeywords={class, export, boolean, throw, implements, import, this},
  ndkeywordstyle=\color{darkgray}\bfseries,
  identifierstyle=\color{black},
  sensitive=false,
  comment=[l]{//},
  morecomment=[s]{/*}{*/},
  commentstyle=\color{purple}\ttfamily,
  stringstyle=\color{red}\ttfamily,
  morestring=[b]',
  morestring=[b]",
  mathescape=true
}

\lstdefinelanguage{yaml}{
  keywords={true,false,null,y,n},
  keywordstyle=\color{darkgray}\bfseries,
  ndkeywords={},
  ndkeywordstyle=\color{black}\bfseries,
  identifierstyle=\color{black},
  sensitive=false,
  comment=[l]{\#},
  morecomment=[s]{/*}{*/},
  commentstyle=\color{purple}\ttfamily,
  stringstyle=\color{blue}\ttfamily,
  morestring=[b]',
  morestring=[b]",
  postbreak=\mbox{\textbf{$\hookrightarrow$}\space},
}
\newboolean{showcomments}
\setboolean{showcomments}{true}         %\setboolean{showcomments}{false} 

\ifthenelse{\boolean{showcomments}}
  {\newcommand{\nb}[2]{
  \fbox{\bfseries\sffamily\scriptsize#1}
     {\sf\small$\blacktriangleright$\textit{\textcolor{red}{#2}}$\blacktriangleleft$}
   }
  }
  {\newcommand{\nb}[2]{}
   
  }

\newcommand{\simsymbol}[1]{\faCog $\,$}
\newcommand{\emusymbol}[1]{\faServer $\,$}
\newcommand{\pstsymbol}[1]{\faPlug $\,$}
\newcommand{\nsimsymbol}[1]{\phantom{\simsymbol{}}}
\newcommand{\nemusymbol}[1]{\phantom{\emusymbol{}}}
\newcommand{\npstsymbol}[1]{\phantom{\pstsymbol{}}}

\newcommand{\toolname}[1]{LiDiTE}

\ifCLASSOPTIONcompsoc
  \usepackage[nocompress]{cite}
\else
  \usepackage{cite}
\fi
\ifCLASSINFOpdf
\else
\fi
\begin{document}
%\title{On the Creation of Featherweight Digital Twins}
\title{\toolname{}: a Full-Fledged and Featherweight Digital Twin Framework}

\author{Enrico~Russo,
      Gabriele~Costa, Giacomo~Longo, Alessandro~Armando, and Alessio~Merlo % <-this % stops a space
   %\thanks{E. Russo, A. Armando, and A. Merlo are with the  Department of Informatics, Bioengineering, Robotics, and Systems Engineering, ..., ITA (email: enrico.russo@unige.it; alessandro.armando@unige.it; alessio.merlo@unige.it)}
  %\thanks{G. Costa is with , ..., ITA (email: gabriele.costa@imtlucca.it)}
  
\IEEEcompsocitemizethanks{\IEEEcompsocthanksitem E. Russo (Corresponding author), G.Longo, A. Armando, and A. Merlo are with the  Department of Informatics, Bioengineering, Robotics, and Systems Engineering (DIBRIS), University of Genoa, Italy.
Email: \{name.surname\}@dibris.unige.it.\protect
%\IEEEcompsocthanksitem G. Longo is with the University of Rome ``Sapienza'' and University of Genoa. Email: giacomo.longo@uniroma1.it.
\IEEEcompsocthanksitem G. Costa is with the Institute for Advanced Studies (IMT), Lucca, Italy. Email: gabriele.costa@imtlucca.it.}% <-this % stops a space
}


\markboth{IEEE Transactions on Dependable and Secure Computing}%
{E. Russo \MakeLowercase{\textit{et al.}}: \toolname{}: a Full-Fledged and Featherweight Digital Twin Framework}



\IEEEtitleabstractindextext{%
\begin{abstract}
The rising of the Cyber-Physical System (CPS) and the Industry 4.0 paradigms demands the design and the implementation of Digital Twin Frameworks (DTFs) that may support the quick build of reliable Digital Twins (DTs) for experimental and testing purposes. 
Most of the current DTF proposals allow generating DTs at a good pace but affect generality, scalability, portability, and completeness. 
As a consequence, current DTF are mostly domain-specific and hardly span several application domains (e.g., from simple IoT deployments to the modeling of complex critical infrastructures). 
Furthermore, the generated DTs often requires a high amount of computational resource to run.\\ 
In this paper, we present \toolname{}, a novel DTF that overcomes the previous limitations by, on the one hand, supporting the building of general-purpose DTs at a fine-grained level, but, on the other hand, with a reduced resource footprint w.r.t. the current state of the art. 
We show the characteristics of the \toolname{} by building the DT of a complex and real critical infrastructure (i.e., the Smart Poligeneration Microgrid of the Savona Campus) and evaluating its resource consumption. 
The source code of \toolname{}, as well as the experimental dataset, is publicly available.
\end{abstract}

%OLD ABSTRACT
%Digital twins are receiving more and more attention.
%Their promise to provide a safe and realistic testing environment for running experiments that would not be possible on a production system.
%The main limitation to a wide adoption of digital twins is that they usually require extremely high computational resources.

%In this paper we present a novel approach for the creation of lightweight digital twins.
%Our main goal is to implement digital twins that both $(i)$ have minimal hardware requirements and $(ii)$ scale on arbitrarily complex infrastructures.
%Moreover, here we present a case study infrastructure to assess our methodology.

\begin{IEEEkeywords}
Digital Twin Framework, Digital Twin, Reference Model, Cyber-physical systems, Digital simulation
\end{IEEEkeywords}}

\maketitle

% is not selected <OR> if conference mode is selected - because compsoc
% conference papers position the abstract like regular (non-compsoc)
\IEEEdisplaynontitleabstractindextext

% \fi
\IEEEpeerreviewmaketitle

\input{intro}
\input{model}
\input{approach}
\input{demo}
\input{related}
\input{conclusion}

\bibliographystyle{IEEEtran}
\bibliography{IEEEabrv,Bibliography}

\end{document}

%% file: intro.tex
\section{Introduction}
\label{sec:introduction}

%A Digital Twin (DT) is a virtual replica of some physical assets, which is built to support tests and experiments (like, e.g., process optimization, predictive maintenance,  and what-if scenarios) that cannot be carried out on the actual system, as they would affect its functionality\ernote{rivedere}. 
In the last years, many authors have put forward different definitions of \emph{Digital Twin} (DT)~\cite{Fuller2020}, also in a relationship with a wide range of applications~\cite{Wright2020}.
Although a terminological consensus is still under development, most authors agree that this concept revolves around creating a virtual replica of a physical asset or system to support tests and experiments (like, e.g., process optimization, predictive maintenance, and what-if scenarios).

In the past, DTs have been usually built from scratch and were strongly tied with the system they replicate.
Nonetheless, as the popularity of DTs has grown in the last years - mostly due to the rise of complex infrastructures, such as those implementing Cyber-Physical Systems (CPS) or the Industry 4.0 paradigm - the demand for rapid development of complex, reliable and scalable DTs increased significantly.  
Digital Twin frameworks (DTF) have been proposed and implemented to deal with such requests. 
A DTF is a software tool aimed at supporting the creation of DTs by automatizing some operations and providing pre-configured virtual components. 

\begin{figure*}[h]
    \centering
    \includegraphics[width=0.9\textwidth]{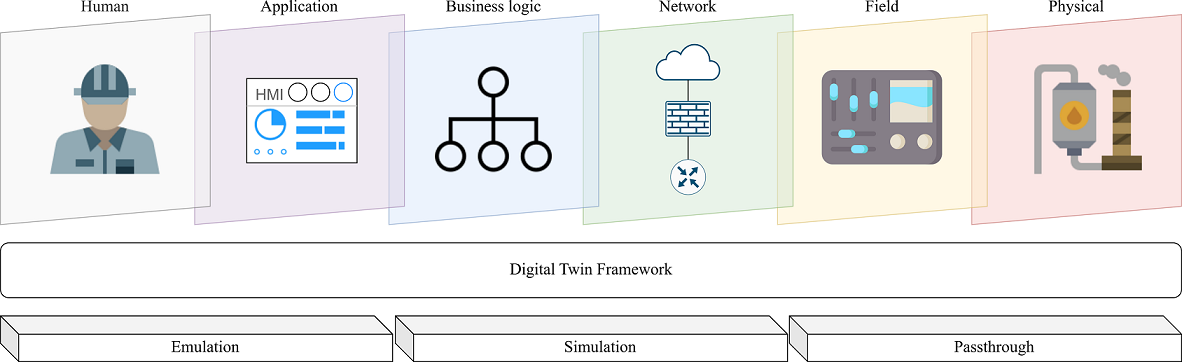}
    \caption{Digital twin framework reference model.}
    \label{fig:overview}
\end{figure*}

Each DTF is based on one reference model, i.e., a general-purpose logical architecture that abstracts some specific parts of a real system.
Consequently, the DTs built through the DTF are instances of (part of) the reference model. 
A reference model should contain, at least, a number of elements abstracting users and components of the real infrastructure and rely on some technologies that allow replicating or mimicking the system's behavior.  
To clarify, a very general reference model is depicted in Figure \ref{fig:overview}. 
 
Here, the elements of the real infrastructures are organized into six layers, namely: 

\begin{itemize}
    \item \textsl{Physical}: physical processes controlled through sensors and actuators.
    \item Field: programmable logic controllers (PLC) and other devices controlling one or more physical element.
    \item \textsl{Network}: the ICT connectivity layer.
    \item \textsl{Business logic}: rules driving the infrastructure behavior.
    \item \textsl{Application}: high-level programs and services.
    \item \textsl{Human}: people operating inside the infrastructure.
\end{itemize}

It is worth noticing that layers are not isolated, but rather they overlap.
The reason is that real systems, and thus their DTs, usually include a complex interplay logic involving different layers.
For instance, think of an operator switching her laptop on. 
Such a simple action has some effects at the physical level  (e.g.,  in terms of energy consumption) and at the network level (e.g.,  in terms of generated traffic). 
This interplay can be referred to as \emph{inter-layers interactions}.

Concerning the replication of the system's behaviors, there exist three types of technologies, namely:

\noindent    
\textbf{Emulation}, i.e., technologies for the creation of virtual replicas of ICT systems such as networks, operating systems, and services. Technologies belonging to this category are, for instance, hypervisors and Linux containers (LXC).

\noindent
\textbf{Simulation}, i.e., technologies abstractly modeling the features of interest of a real system while neglecting implementation aspects. 
Often, simulators are based on some mathematical specification of the behavior of the real system, e.g., given in terms of differential equations or finite state machines.

\noindent
\textbf{Pass-through}, i.e., technologies relying on a real, physical implementation that is directly plugged in the DT. 
This category includes, for instance, hardware, remote services, and human operators. 
Moreover, some technologies are needed to support pass-through integration, e.g., think of virtual private networks (VPN), remote desktop applications, and remote terminals.

The reference model affects the complexity of the DTF and the demand for computational resources of the corresponding DTs.
%As a consequence, most of current DTFs rely on simpler reference models (w.r.t. the model in Figure \ref{fig:overview}) that contains a subset of layers and technologies only (see Section \ref{sec:related} for more information), to find a reasonable balancing between completeness of the replica, and both the complexity to build the DT and the demand of computational resources to execute the DT.\ernote{non mi è chiarissima questa frase} 
As a consequence, most of current DTFs rely on simpler reference models (w.r.t. the model in Figure \ref{fig:overview}) which contain only a subset of layers and technologies (see Section \ref{sec:related} for more information). 
Such simplification is often induced by the act of balancing between completeness of the replica and its computational and development complexity.
We argue that an open research challenge is to build a DTF that may go beyond the need of such a trade-off, and that could be, at the same time:

\begin{enumerate}
    \item \textbf{General Purpose}, i.e., it should rely on a very general reference model, thereby allowing to build DTs of several kinds of physical assets (from simple CPS scenarios to complex critical infrastructures).
    \item \textbf{Expressive}, i.e., the DTF should allow to model all layers, elements, and technologies of interest.
    \item \textbf{Extensible} i.e., the DTF (and the corresponding reference model) should be enhanced by adding new components. 
    \item \textbf{Affordable}, i.e., the DTF should rely as much as possible on existing and off-the-shelf solutions. 
    \item \textbf{Lightweight}, the generated DTs must require a limited amount of computational resources to work properly. 
\end{enumerate}

To this aim, in this paper, we present the design and the implementation of a novel DTF, called  \textbf{LiDiTE}, which satisfies all previous properties.  
Besides the compliance w.r.t. the model in Figure \ref{fig:overview},  the inspiring principle of LiDiTE is to combine existing and new, ad-hoc technologies to require minimal computational resources. 
Moreover, we apply LiDiTe to implement a DT of a real, complex facility, i.e., the \textsl{Savona  Polygeneration  Microgrid}  (SPM)~\cite{spm}. 
Our experimental results show that, despite the low demand for computational resources, the performance of our DT implementation overlaps with that of the original SPM infrastructure. 

The rest of the paper is structured as follows. Section~\ref{sec:model} introduces our DTF model which we implement in Section~\ref{sec:approach}.
In Section~\ref{sec:demo} we demonstrate our implementation by applying it to the SPM use case. 
Finally, in Section~\ref{sec:related} we revise the related work and we conclude the paper in Section~\ref{sec:conclusion}.

%% file: model.tex
\section{A Reference Model for Digital Twins}
\label{sec:model}

In this section we will describe in details the reference model of our DTF.

\subsection{Overview}
\label{sec:model_overview}

\begin{figure*}[t]
    \centering
    \includegraphics[width=0.95\textwidth]{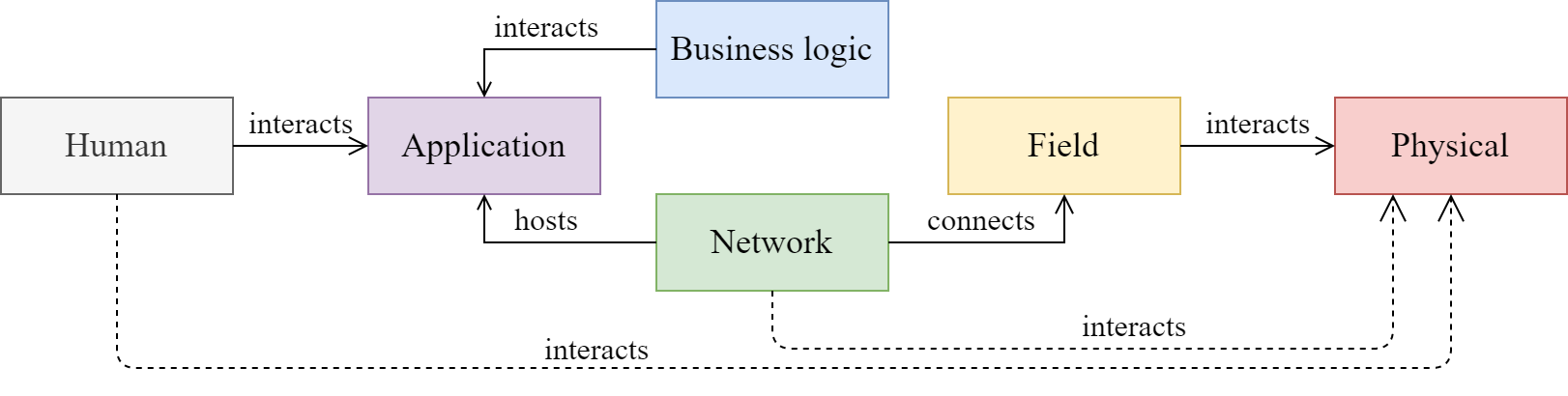}
    \caption{\toolname{} architecture and inter-layers interactions.}
    \label{fig:lidite_model}
\end{figure*}

In principle, building an accurate copy of a physical system requires identifying and replicating all of its components and functionalities and their interplay.
Needless to say, replicating every single aspect of the physical environment is highly complex (likely infeasible) and, in general, out of scope w.r.t. the purposes of a DT.
Thus, we consider a reference model based on the relationships highlighted in Figure~\ref{fig:lidite_model}.

The main idea behind this model is to prioritize interactions that we consider more frequently relevant for common usages of DTs.
Most of these interactions are implemented by leveraging some existing technologies (solid arrows).
Instead, other interactions (dashed arrows) have been implemented \emph{ex nihilo} due to both their relevance and the absence of specific solutions.
In particular, we considered the interactions of the physical environment with both human agents and network infrastructures.
Full details about these aspects are provided in Section~\ref{sec:approach}.

As previously stated, our reference model follows the separation of concerns principle by identifying six structural layers.
Our layers mimic those defined in some standard reference models, such as Purdue Enterprise Reference Architecture (PERA)~\cite{williams1998purdue} or Reference Architectural Model Industrie (RAMI)~\cite{hankel2015reference}. 
Nevertheless, our layer interaction model is \emph{open}, i.e., no specific constraints limit the possible interactions between the elements of two layers.
In other words, a DTF relying on this model can also extend it with further interactions, e.g., for representing a specific behavior of interest.
All the interactions that involve distinct layers are called \emph{inter-layer interactions}.
For the time being, we only focused on the inter-layer interactions highlighted in Figure~\ref{fig:lidite_model}.

In the following sections, we describe each layer and the components/technologies they integrate. 
Finally, in Section~\ref{sec:model_inter}, we detail the inter-layer interactions introduced above. 

\subsection{Physical layer}
\label{sec:model_physical}
This layer concerns the physical environment, phenomena, and objects that somehow affect the infrastructure.
For instance, details such as weather conditions, sun position, radiance, air pollution, etc., belong to this layer.
Although, in principle, myriads of physical elements might be involved in the design of a DT, in practice, only a few of them typically play an active role. 
We assume that interaction through proper sensors/actuators is viable for each physical state of interest.
Briefly, sensors read a value from the physical environment (e.g., a temperature), while actuators modify it (e.g., a heater).
In other words, we assume sensors/actuators to be the interface between the physical world and any other element of a DT.

\subsection{Field layer}
\label{sec:model_field}
The field layer includes devices that handle data from and to the physical layer.
A field device may collect data from sensors, update its internal state, and trigger actuators.
It takes decisions based upon a custom program, i.e., its control logic.
By leveraging direct connection with the network layer, field device functionalities are also accessible from other layers.
For example, consider a field device controlling a valve (physical layer) that regulates a pipe stream.
A remote, networked system can query the field device about the current stream, depending on the valve status.
Also, the same system can request the reduce the stream, which triggers the field device control logic. 
As a consequence of the control logic computation, the field device may send the closure command to the valve actuator.

Among the elements of the field layer, PLCs and IoT devices are arguably the most common and representative.
The reason is that PLCs provide the core functionalities of the legacy industrial automation, while IoT devices are essential for implementing the Industry 4.0 paradigm.

\subsection{Network layer}
\label{sec:model_network}

This layer includes all the elements contributing to the network infrastructure.
Each network infrastructure consists of nodes, e.g., hosts, routers, firewalls, and the networks connecting them.
A host node can be, for example, a server, a personal computer, or a networked PLC.
Networks provide connectivity to nodes and enable their communication, e.g., through standard protocols.
Typically, enterprise networks include several segments, i.e., sub-networks, devoted to a specific task, e.g., a laboratory network or class of nodes, e.g., IoT devices.
Network devices, e.g., routers or firewalls, are the nodes enabling the overall internetworking. 

\subsection{Business logic layer}
\label{sec:model_business}

The business logic layer involves the processes driving the infrastructure behavior.
Industrial processes, optimization heuristics, centralized and distributed control logic belong to this layer.
In general, the business logic of a complex infrastructure may involve many processes.
Each process is executed by one or more agents, e.g., machines and human beings.
This work mainly focuses on the business logic of industrial control systems (ICS) such as SCADA.
Such processes are typically in charge of ensuring that the infrastructure behavior is optimal and safe.
In particular, a command and control application acquires and aggregates real-time data from the field.
Also, according to the specific business logic, commands are sent back to field devices.
This task may also involve human actors using the SCADA HMI application.

\subsection{Application layer}
\label{sec:model_application}

This layer refers to the software elements, such as applications and operating systems. 
Resident applications may significantly vary between different infrastructures.
In general, complex infrastructures host many applications that coexist, e.g., think of the applications running in parallel on a single personal computer.
Some applications are related to other layers.
For instance, we already discussed above how SCADA systems manage the business logic of the infrastructure.
Also, most applications of interest are the source/destination of network traffic.

\subsection{Human layer}
\label{sec:model_human}

The human layer includes people that exist inside the infrastructure.
Human beings can operate on a real infrastructure in many (sometimes unexpected) ways.
In general, humans can interact with applications (e.g., by clicking on a button), networks (e.g., by plugging a network cable), business logic (e.g., by changing an optimization process), field devices (e.g., by using the controls of a PLC) and physical environment (e.g., by manually closing a valve).
Although all of these interactions may be relevant, we focus on the application and physical layers here.
The reason is that in many cases, human agents are expected to interact with the infrastructure through one of these two layers.
For instance, a SCADA operator sitting in a control room mainly interacts with the SCADA HMI and, perhaps, some other applications residing on her computer.
Furthermore, people moving inside the infrastructure are likely to induce some physical effects, e.g., think of a motion sensor.

\subsection{Inter-layer interactions}
\label{sec:model_inter}

An inter-layer interaction occurs when a components belonging to different layers influence each other.
Figure~\ref{fig:lidite_model} shows the inter-layer interactions currently supported by \toolname{}.
Since multiple types exist, we use arrows to denote the overall family of interactions relating two layers.
We remark that arrows are merely symbolic as actual interactions are typically bidirectional.
For instance, consider the relationship between Human and Application.
A human operating through a user interface is involved in a bidirectional interaction where she provides inputs and reads outputs.
Moreover, it is worth noticing that the effects of an inter-layer interaction can propagate through multiple layers, by following a relationships chain.
For example, a human agent interacting with a SCADA HMI (Application layer), can influence the ICS business logic and result in executing a command for the field devices acting on the physical environment.
In the next section, we provide the implementation details about the currently supported inter-layer interactions.

%% file: approach.tex
\section{\toolname{}}
\label{sec:approach}

In this section we present our DTF called \emph{Lightweight Digital Twin Environment} (\toolname{}).

\subsection{Overview}
\label{sec:overview}

\toolname{} includes all the six layers discussed above, and, in particular, they are implemented using different combinations of the three technological pillars, i.e., emulation, simulation, and passthrough.
Passthrough is supported at every layer to allow for \emph{plug-and-play} integration with external elements.
Instead, simulation and emulation technologies are used at different layers to implement specific features of a DT.
Full details about the implementation of both each layer and inter-layer interactions are given below.

\subsection{Physical layer}
\label{sec:physical}
As we discussed in Section~\ref{sec:model_physical}, the Physical layer refers to the physical elements that exist inside the infrastructure.
In \toolname{}, such physical assets can be either simulated or directly integrated (passthrough).
A passthrough configuration allows the DT to interact with real world sensors and actuators.
For example, a PLC (Field layer) can read the temperature from a real sensor.
\toolname{} supports the above connections by leveraging on the physical interfaces provided by the host running the DT, e.g., serial or USB ports.

Physical components simulation requires more attention. 
As a matter of fact, \toolname{} mainstream simulation frameworks, e.g., Mathworks Simulink~\cite{simulink21}, can be integrated through software emulation (see Section~\ref{sec:applayer}).
Nevertheless, full-fledged simulators are typically not implemented to be lightweight\footnote{https://mathworks.com/company/newsletters/articles/improving-simulation-performance-in-simulink.html}, which would result in poor scalability.
For this reason, \toolname{} includes a lightweight simulation module.
The module supports three simulation methods which we describe below.

\paragraph*{\textbf{Interpolation}}

This method simulates the behavior of a system by interpolating a given dataset, e.g., historical sequences.
The dataset is a finite mapping between input $x_i$ and output $y_i$ values.
Given a new input $x$, this method retrieves the closest elements $x_1, \ldots, x_k$ from the dataset inputs and applies an interpolation function $f$ to the corresponding outputs $y_1, \ldots, y_k$.
By default, \toolname{} applies the linear interpolation function to the values $y_1, y_2$ such that $[x_1, x_2]$ is the smallest interval in the dataset inputs with $x \in [x_1,x_2]$.
Alternative interpolation functions can also be defined through the Script method (see below).

\paragraph*{\textbf{ODE simulation}}

Ordinary differential equations (ODEs) are often used to describe dynamic systems.\footnote{We refer the interested reader to~\cite{Sontag98control}.}
With this method, a system is modeled through an ODE, given in the canonical form
$$
\dot{x}(t) = A(t)x(t) + B(t)u(t)
$$
\noindent
where $A(t) \in \mathbb{R}^{n \times n}$, $B(t) \in \mathbb{R}^{n \times m}$, $x(t) \in \mathbb{R}^{n}$, and $u(t) \in \mathbb{R}^{m}$. 
Briefly, the system's evolution  at time $t$ is given by a linear combination of its current state $x(t)$ and its current input $u(t)$.
In this context, matrices $A$ and $B$ express the dependency of the next state w.r.t. the current state ($A$) and the current inputs ($B$, respectively).
In \toolname{}, ODEs can be specified through JSON files.

\paragraph*{\textbf{Scripting}}
The third simulation mechanism amounts to executing arbitrary JavaScript code.
In this way, one can develop any simulation logic of interest, even when neither historical data nor an ODE is available.
Interestingly, this also enables the integration of the simulator with external resources, e.g., a remote service.

\subsection{Field layer}
\label{sec:impl-field}
In \toolname{}, field devices can be implemented via emulation and passthrough.
This is achieved by means of Eclipse Ditto~\cite{eclipse21ditto} and OpenPLC~\cite{openplc}, respectively.

\paragraph*{\textbf{Eclipse Ditto}}

Ditto is a mainstream software for creating DTs of field devices.
Briefly, Ditto connects different physical devices, mirrors them, and enables a single point of access through unified APIs and protocols.
In the above configuration, it allows \toolname{} to implement the passthrough and integrate the DT with existing devices that support standard IoT transport protocols~\cite{Naik2017}, e.g., AMQP or MQTT.
Moreover, Ditto is used for implementing field device emulation. 
In particular, Ditto permits the creation of virtual replicas of field devices that connects with objects at the physical layer (see~\ref{sec:physical}).

\paragraph*{\textbf{OpenPLC}}

OpenPLC is an open-source solution for PLC virtualization.
In general, OpenPLC executes programs that comply with the IEC 61131-3 standard~\cite{Ramanathan2014TheI6}, e.g., ladder logic programs, as industrial PLCs do.
For instance, a virtual PLC can be configured to act as a Modbus~\cite{swales1999open} Slave or as a DNP3~\cite{curtis2005dnp3} outstation to emulate, e.g., a field device managed by a remote SCADA server.
Moreover, OpenPLC can be used to implement passthrough.
In particular, this is enabled when a virtual PLC is configured to act as a Modbus Master that works as a proxy for a real PLC being the Modbus slave.

\subsection{Network layer}
\label{sec:impl-network}
In \toolname{} networks and network behavior are obtained by means of emulation, simulation, and passthrough technologies.
Their implementation relies on the Docker networking system and software for creating OSI layer 2 Virtual Private Network, namely OpenVPN~\cite{openvpn21}.
The Docker networking system is configured with a plugin that
leverages the Software-Defined Networking (SDN) architecture and the OpenFlow protocol~\cite{rfc7426}.
In particular, the above plugin manages an SDN infrastructure deployed with Faucet~\cite{Bailey16Faucet} as the OpenFlow controller and Open vSwitch~\cite{188960} for the virtual switches connecting DT nodes.

By default, \toolname{} provides three network modes, i.e, \emph{flat}, \emph{routed}, and \emph{nat}.
Flat networks provide connectivity between connected nodes only and can be used to emulate the internal segments of, e.g., an enterprise network.
Routed and nat networks provision connected nodes with a gateway through the host platform, thus enabling a direct link between the nodes and the host.
Moreover, by applying Network Address Translation, nat networks masquerade addresses of nodes with the host's one.
Typically, they are used to implement \emph{management} and \emph{host-through} networks.
A management network allows DT owners to issue direct messages to some nodes, e.g., for running commands and monitoring purposes.
Although a management network is not mandatory, it is actually needed in several cases, e.g., to trigger critical events, such as a black-out or a system failure.
Instead, host-through networks allow implementing segments accessing the host resources, e.g., its services or hardware.
Host-through networks are a fundamental building block for network-based passthrough integration.
As a matter of fact, they allow DT nodes to address external resources, e.g., remote, Internet-connected systems, by leveraging the host connectivity. 

\toolname{} also supports network and node passthrough integration, i.e., any external networks or devices can be directly connected to any DT network segment.
Network passthrough is obtained by configuring the Faucet service to manage external, physical, or virtual switches supporting the OpenFlow protocol. 
Instead, node passthrough can be implemented using OpenVPN.
The configuration consists of a one-to-one mapping between the OpenVPN server network interfaces and the passthrough-enabled DT networks.

Beyond network connectivity, network devices must be appropriately configured to implement internetworking.
The above devices are emulated through multihomed containers and leveraging Docker images for running their operating system.
As we detail in Section~\ref{sec:applayer} for emulated applications, \toolname{} can use containers for Linux based operating systems, e.g., OpenWrt~\cite{openwrt21}, or virtual machines for the devices that are provided in the form of virtual appliances, e.g., OPNsense~\cite{opnsense21} or the virtualized form of firewalls provided by major vendors. 

\begin{figure}
    \centering
    \includegraphics[width=\columnwidth]{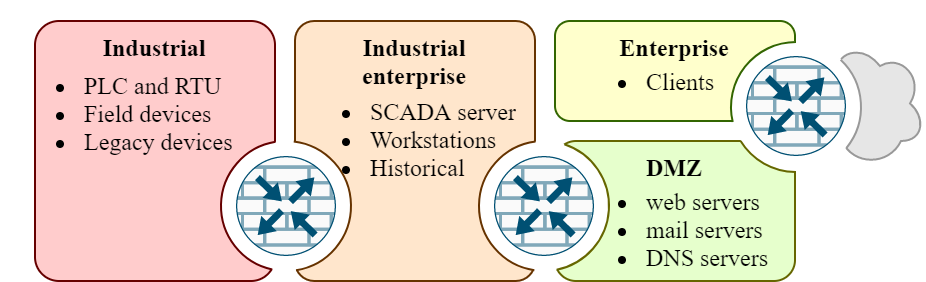}
    \caption{The network chassis layout.}
    \label{fig:schema}
\end{figure}

Finally, standard and frequently used networks can be defined and reused in \toolname{}.
A reusable network topology goes under the name of \emph{Network Chassis}.
For instance, The network chassis of Figure~\ref{fig:schema} follows a recurrent ICS network architecture featuring the segmentation model proposed in the ISA/IEC 62443 standard~\cite{international2010iec}.
Briefly, it consists of three firewalls enforcing four security zones assigned to $(i)$ publicly exposed IT services, i.e., DMZ, $(ii)$ internal IT services, i.e., Enterprise, $(iii)$ services for the industrial automation, i.e., Industrial enterprise, and $(iv)$ field devices, i.e., Industrial.
Our network chassis also supports connectivity with the real or a simulated Internet through a fourth emulated router working as an Internet Service Provider (ISP).
Finally, the support of an emulated DNS server and dynamic routing protocols allow DT designers to adapt a predefined network chassis to represent the domain names and public and private addressing.

\subsection{Application layer}
\label{sec:applayer}
\toolname{} hosts applications by relying on the passthrough or emulation technologies.
As previously detailed in Section~\ref{sec:impl-network}, network passthrough allows for plugging a real host into a DT network which, clearly, also entails passthrough integration of all the applications residing there.
Nevertheless, a single application passthrough might be necessary, e.g., to deploy an existing service inside a context relevant for a specific DT.
For instance, in \toolname{} this is achieved by using \emph{reverse proxies}.
For the time being, \toolname{} supports reverse proxy by using NGINX~\cite{reese2008nginx}.
Briefly, NGINX embeds a reverse proxy module that can be configured for both TCP and UDP traffic.
For example, we can mount the NGINX module on a DT host and configure it to mirror an existing web portal.

For the emulation, \toolname{} resorts to Linux Containers and, in particular, to Docker.
The main reason is that, since Linux containers only carry small runtime environments, the scale well and better match the lightweight requirements of \toolname{}.
Still, other emulators, e.g., VM hypervisors, are supported.
For instance, \toolname{} integrates KVM~\cite{kvm} through a wrapper container.
This solution allows the DT to emulate non-Linux applications, e.g., Microsoft Windows ones.

In addition, by means of Docker Compose~\cite{compose21}, \toolname{} can integrate out-of-the-box applications consisting of several modules.
For instance, Eclipse Ditto (see above) is imported by \toolname{} in a way that its dependencies, e.g., Java runtime or the Mongo database, are already satisfied and pre-configured.

\subsection{Business logic and Human layers}
\label{sec:controlhuman_layer}
Business Process Modelling and Notation (BPMN)~\cite{white2004introduction} is the default process definition language supported by \toolname{}.
Being widely adopted, BPMN is supported by many tools, and \toolname{} relies on the Camunda platform~\cite{camunda2021}, i.e., a mainstream, open-source software for designing and executing BPMN processes.
Briefly, in Camunda, each task is associated with a corresponding script, e.g., implemented in Groovy, which is launched whenever the task node is visited.
In \toolname{} an instance of Camunda is available at the Application layer for enabling business logic execution through emulation.
Nevertheless, due to their expressive power, we also exploit BPMN processes for simulating human behavior.
For instance, in this way, Camunda can be instructed to simulate users who browse the web, e.g., by visiting URLs from a predefined list.
This is achieved by coding REST API invocations as part of the task scripts.

Finally, business logic and human agents integration can happen via passthrough.
In particular, human agents can operate through the DT interaction points, e.g., applications and physical elements.
Also, external business logic applications can be connected by means of application passthrough technologies (see above).

\subsection{Inter-layer interactions}
\label{sec:interlayer}

\begin{figure}
    \centering
    \includegraphics[width=\columnwidth]{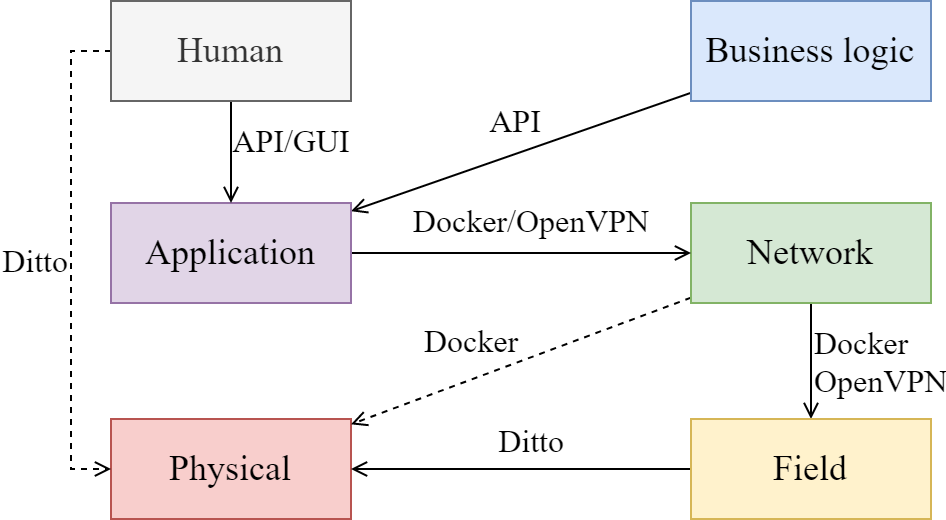}
    \caption{Inter-layer interactions implemented in \toolname{}.}
    \label{fig:interimpl}
\end{figure}

In this section, we provide the reader with the implementation details about the inter-layer interactions currently supported in \toolname{}.
Figure~\ref{fig:interimpl} schematically depicts them.
We first observe that two families of interactions exist, i.e., those inherited from the specific technologies adopted in \toolname{} (solid arrows) and those developed on purpose (dashed arrows).

Schematically, the inherited interactions are the following.
Simulated humans trigger APIs via Camunda scripts, while real humans directly use the graphical user interfaces of the applications. 
Similar to simulated humans, the business logic processes only operate through APIs.
Emulated applications use Docker interfaces to interact with the network hosting them, while passthrough applications can resort to both open Docker and OpenVPN.
The same goes with emulated and passthrough field devices connected to the network infrastructure.
Finally, Eclipse Ditto APIs are used for the communications between field devices and both emulated and passthrough physical elements.

The current version of \toolname{} includes two further inter-layer interaction mechanisms that we implemented as follows.
To enable interactions between simulated humans and simulated physical elements, Ditto APIs are made callable from Camunda processes.
In this way, simulated humans can manipulate the state of simulated physical systems in the same way as field devices.
Instead, Docker APIs enable, among others, the listing, creation, power on/off, and scale of containers simulating network nodes.
Again, Camunda can interact with such APIs to implement the effects of emulated network nodes on simulated physical elements, e.g., the increase of the power consumption, and vice versa, e.g., the impact of a power outage.

%% file: demo.tex
\section{\toolname{} Demo}
\label{sec:demo}

We provide a demonstration of \toolname{} applied to a real use case.
\toolname{} is available as a free, open source software on GitHub~\footnote{\url{\repo}}.
%More precisely, the version used for this article is \toolname{} v1.0.0~\ernote{ref a zenodo}.
All the material needed to replicate the use case DT as well as the experiments described in Section~\ref{sec:demo-evaluation} are also available in the GitHub repository.

\subsection{Use Case}
\label{sec:usecase}

\begin{figure}[t]
     \centering
     \begin{overpic}[width=\columnwidth]{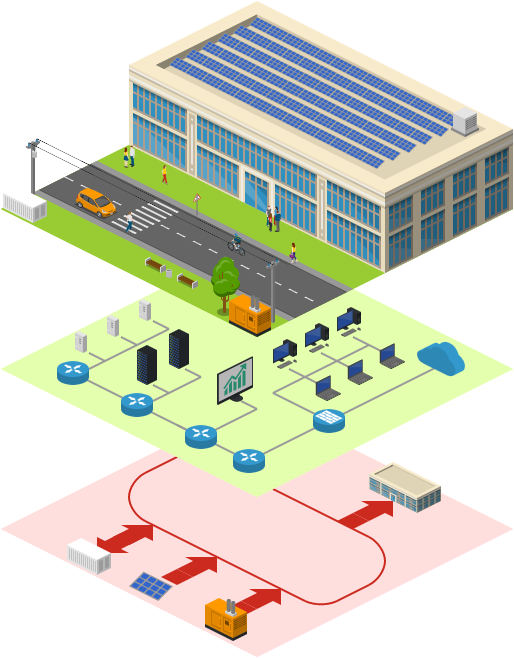}
     \put (60,1) {\small (icograms.com)}
     \end{overpic}
     \caption{A schematic representation of the SPM infrastructures.}
     \label{fig:spm}
\end{figure}

The Smart Polygeneration Microgrid is a power generation plant deployed within the Savona campus of the University of Genoa.
The campus hosts a number of facilities for university students and staff, including a gym and a tennis court.
Classrooms, laboratories, and offices are located inside five buildings.
Moreover, a sixth building contains the library and some study rooms.
In general, the SPM builds on top of three distinct, yet interconnected, infrastructures, i.e., the university campus, the ICT network, and the energy system (see Figure~\ref{fig:spm}).
These infrastructures host several subsystems.
Below, we provide a detailed description of those that are relevant to this paper.

\subsubsection{Power plant}

The microgrid structure is that of a power ring connecting all the subsystems.
The main power generation subsystem amounts to a solar plant installed on the roof of building B. 
Solar panels can output up to 80.64 kW~\cite{Bracco17smart}, depending on environmental factors, such as season, daytime, and weather conditions.
If needed, a Capstone\textsuperscript{\textregistered} C65~\cite[\S 4.1]{bracco-gas-turbine} gas turbine can be activated to integrate the energy production.
The turbine is a co-generative system that combines combustion and heat, producing 65 kW.
Furthermore, energy storage is recharged to accumulate energy surplus and discharged to cover the energy deficit.
The overall energy demand amounts to the sum of demands of each building, which depend on the local energy consumption.
Finally, the SPM drains from the public energy distribution network in case of excess demand.
%The overall structure is depicted in Figure~\ref{fig:physical}.

% \begin{figure}
%     \centering
%     \includegraphics[width=\columnwidth]{fig/physical.png}
%     \caption{The SPM power ring}
%     \label{fig:physical}
% \end{figure}

\subsubsection{On field controllers}

Each subsystem is under the control of a field device.
Field devices are connected to the central SCADA system (see Section~\ref{sec:scada}).
Field controllers feed the SCADA server with data from their own subsystem, e.g., the currently produced power, and receive commands to be applied, e.g., to switch on and off the subsystem.
Communications are implemented in two ways.
Buildings cabinets are directly connected with the SCADA server through Modbus/TCP~\cite{swales1999open}.
Instead, storage, turbine, and solar controllers interact using the Publish/Subscribe pattern.
The connection occurs through a message broker server which handles the message queue between the field devices and the SCADA server.
The SCADA server relies on REST APIs to query the broker.

\subsubsection{Campus network}

% \begin{figure*}
%     \centering
%     \includegraphics[width=\textwidth]{fig/network.png}
%     \caption{ICT network scheme}
%     \label{fig:network}
% \end{figure*}

\begin{figure}
    \centering
    \includegraphics[width=\columnwidth]{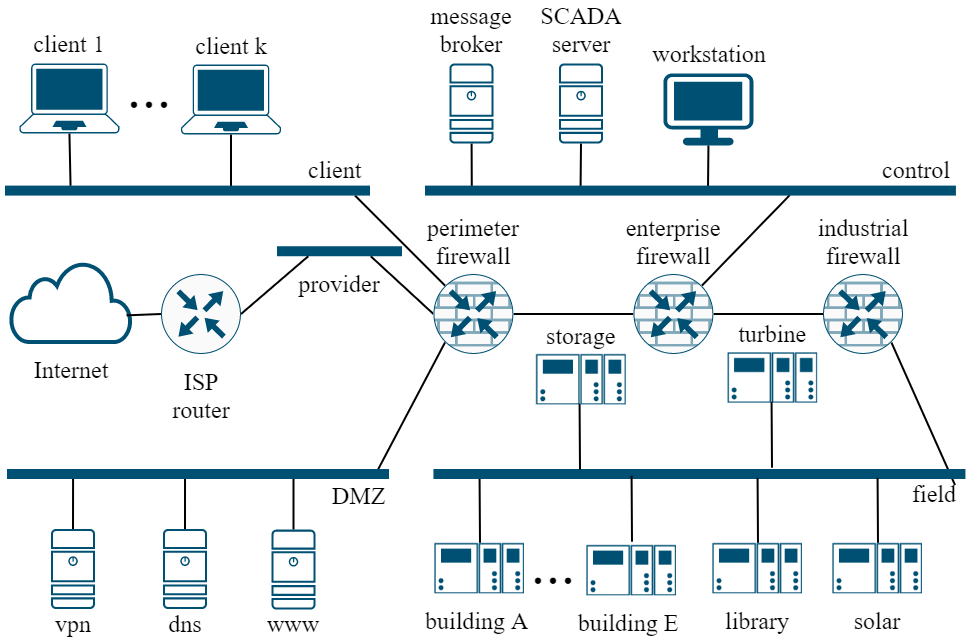}
    \caption{ICT network scheme}
    \label{fig:network}
\end{figure}

The campus ICT infrastructure consists of a segmented enterprise network (see Figure~\ref{fig:network}).
The four main segments are called \emph{client}, \emph{DMZ}, \emph{control}, and \emph{field}.
The client network hosts personal devices belonging to staff and students, and it is accessible throughout the entire campus.
Concerning the campus servers, the DMZ network  is located inside the server room, and it is meant to expose remotely accessible, public services.
The control network hosts servers and workstations dedicated to the smart grid management.
The field network hosts the field devices.
Finally, the campus is connected to the Internet through an Internet Service Provider (ISP) router.
The overall segmentation is enabled by three firewalls that allow each network to operate under distinct security policies.
For example, a policy enforced on the field firewall denies field devices to browse the Internet and only allows incoming connections from the control network.

\subsubsection{Energy Management System}
\label{sec:ems}

The Energy Management System (EMS) is a centralized, automatic control process that handles the SPM power generation logic.
The goal of the EMS is to employ the power generation subsystems to maximize the SPM efficiency in terms of costs and ecological footprint. 
The EMS monitors the internal demand and the overall power generation to this aim.
If possible, solar energy generation is privileged since it comes at almost no cost and has negligible effects on the environment.
When solar production exceeds the demand, the energy surplus is redirected to energy storage.
To do this, the EMS activates the \emph{charge} mode of the energy storage.
If the energy storage cannot switch to charge mode, e.g., because the full charge has been reached, the surplus is dissipated.
When the solar production is insufficient, the EMS attempts to cover the energy deficit by executing the following steps.
\begin{enumerate}
    \item[1.] The energy storage is set to \emph{discharge} mode.
    \item[2.] If 1. is unfeasible, the gas turbine is started.
    \item[3.] If 2. is unfeasible or uneconomical, energy is bought from the public supplier. 
\end{enumerate}

\subsubsection{SCADA system}
\label{sec:scada}

The SPM is under the scope of a SCADA system.
The SCADA system acquires data from the peripheral subsystems and allows operators to monitor the SPM.
Furthermore, operators can manually instruct the SPM by modifying the state of the system.
The SCADA system runs on the SCADA server (see above), and operators access its interface through a dedicated workstation inside the SPM control room.

\subsubsection{Campus staff and students}

Mainly two categories of people move inside the campus, i.e., staff members and students.
Staff members include SPM operators and teaching and administrative personnel.
Students and staff members can connect to the SPM enterprise network with their client devices, e.g., laptops, mobile, and wearable devices.
The number of people inside the campus varies depending on some factors such as the daytime and the period of the year.

\subsection{Implementation}
\label{sec:demo-impl}

Our DT implementation includes the elements of SPM that we detailed in the previous section.
In particular, we have coupled each of the above elements to a layer of our reference model and used the emulation and simulation technologies supported by \toolname{} for creating their instance in the DT.
For each layer, we discuss below the implementation details using some representative elements as examples.

\subsubsection{Physical layer}
\label{sec:demo-physical}
At this layer, we require to model the processes related to the solar panel subsystem, the gas turbine, and the energy storage.
As detailed in Section~\ref{sec:physical}, we can simulate them using the lightweight module included in \toolname{}.
In the following examples, we detail the methods and configurations we used to model their behaviors.

\begin{example}
\label{ex:interpolation}
Consider the solar panel subsystem of the SPM use case.
Historical data is often used for predicting the expected performance, i.e., the generated power.
Since we have no data about the production of similar plants, we derive it from historical data about solar radiation.
We obtain this through an ancillary sun simulator, discussed in detail in Example~\ref{ex:scripting}.
Given the current time $x$ (date and daytime), the sun simulator returns a radiation value $y$ (in $W/m^2$).
Then, we compute $f_p(y) = y S E$ where $S$ is the panel area (in $m^2$) and $E$ is the panel efficiency (in $\%$).
The terms appearing in $f_p$ are based on the technical specifications of the system.
In particular, the full specifications of the solar plant is provided in~\cite{Bracco17smart}; this includes the overall surface $S$ and the solar modules specifications\footnote{\url{https://raw.githubusercontent.com/CSecLab/LiDiTe/master/config/scriptablesensor/energy-store-1/Sistem_Sonick_ST5_23_620V.pdf}} from which $E$ is taken.
Below we give the JSON syntax for the simulator described above.

Briefly, the code below defines a device called \emph{FDT:solar-panel-1} possessing a property \emph{power} which is calculated by invoking the indicated \emph{getSolarSurfaceInterpolant} function with the \emph{watt-per-msq} value obtained from the latest reported state of another device called \emph{FDT:sun-simulator}.

\begin{lstlisting}[language=JavaScript,basicstyle=\ttfamily\footnotesize]
{"name": "FDT:solar-panel-1", "features": {
  "solar-panel": {
   ...
   "components": [{ "sources": [{
     "thingId":"FDT:sun-simulator",
     "feature":"environment",
     "pointer":"/watt-per-msq"
    }], "callbackName":"getSolarSurfaceInterpolant",
    "name":"power"}]}}}
\end{lstlisting}

Finally, the JavaScript callback function implementing $f_p$ is as follows. 

\begin{lstlisting}[language=JavaScript]
function getSolarSurfaceInterpolant($y$, $S$, $E$) {
  return ($y$) => $y$ * $S$ * $E$; }
\end{lstlisting}
\end{example}

\begin{example}
\label{ex:ode}
Consider the energy storage subsystem.
Following the technical specifications of the manufacturer\footnote{\url{https://raw.githubusercontent.com/CSecLab/LiDiTe/master/config/scriptablesensor/solar-panel-1/ferrania_solis.pdf}}, we model it as follows.
$$
\dot{x}(t) = [0] x(t) + [0.126, -0.14] u(t)
$$
In this case, the current state of the system $x(t)$ amounts to a single value, i.e., the storage charge level.
The charge level variation only depends on the inputs, i.e., $A(t) = [0]$.
In particular, two inputs are given, namely \emph{recharge} and \emph{discharge}.
The JSON specification for the previous ODE is the following.
\begin{lstlisting}[language=JavaScript,basicstyle=\ttfamily\footnotesize]
{"name": "FDT:energy-store-1", "features": {
  "battery-pack": { "type": "systemSimulator",
   ...
   "system": {"A": [[0]], "B": [[0.12667, -0.14]]}}}}
\end{lstlisting}

Also, the gas turbine is modeled with an ODE as follows.
\begin{small}
$$
\dot{x}(t) = \left[ \begin{array}{c c} -0.3076 & 0 \\ 0.0008 & -0.2 \end{array} \right] x(t) + \left[ \begin{array}{c c c} 4750 & 29993 & -0.1 \\ 1 & 45 & 0.2 \end{array} \right] u(t)
$$
\end{small}
Here, $x(t)$ is made up of two components, i.e., the current number of revolutions per minute and the exhaust gas temperature in Celsius degrees.
The system state is influenced by three inputs, i.e., startup valve enabled (ranging in $\{0,1\}$), ignition valve enabled (ranging in $\{0,1\}$), and outside air temperature (ranging in $[-7,39]$\cite{arpal-clima}).
The coefficients of $A(t)$ and $B(t)$ have been calculated to match the manufacturer-provided specifications about turbine performance (in the 50\%-100\% rpm range).
\end{example}

\begin{example}
\label{ex:scripting}
To simulate the sun radiance, we leverage the European Commission's Photovoltaic Geographical Information System (PVGIS)~\cite{pvgis21}.
The service exposes APIs retrieving the solar radiance component for specific coordinates, already corrected according to specified rise and azimuth, at a given date.
A minimal JavaScript implementation of the simulator is given below.

\begin{lstlisting}[language=JavaScript,basicstyle=\ttfamily\footnotesize,mathescape,morekeywords={let}]
const latitude = 44.18, longitude = 8.18
const azimuth = -30, rise = 15, year = 2016
const data = createSolarDataTable()
function createSolarDataTable() {
  let response = RESTUtil.doGet('https://...')
  // parsing and cleaning PVGIS API response
  return cleaned_up_data_table }
function getRadianceAtTime($x$) {
  let record = $\min_{|x - d|}$ data.$d$ s.t. $d \in$ data.keys()
  return record.radiance }
function getCurrentRadiance() {
  let $y$ = getRadianceAtTime(Date.now())
  return $y$ }
\end{lstlisting}

Briefly, the code above declares constants for latitude, longitude, azimuth, and rise of the solar plant.
As previously discussed, these values are taken from~\cite{Bracco17smart}.
Also, the script declares the reference year for data to be retrieved.\footnote{At the time of writing, the most recent records available are from 2016.}
These values are used by the function \verb|createSolarDataTable|, which invokes the REST APIs of PVGIS and generates a table, called \verb|data|, mapping timestamps of the days of the year to the radiance recorded at that time (in 2016).
Then, the function \verb|getRadianceAtTime|, given a timestamp $x$, searches \verb|data| for the record which is temporally closer to $x$, and returns its radiance.
Finally, when invoked, the function \verb|getCurrentRadiance| computes the radiance for the current time.
This operation generates the value $y$ used by the simulator of Example~\ref{ex:interpolation}.
\end{example}

\subsubsection{Field layer}
\label{sec:demo-field}
Some representative elements that we emulate at the Field layer are the power cabinets and the field controllers.
In Example~\ref{ex:cabinet}, we start detailing the implementation of power cabinets.

\begin{example}
\label{ex:cabinet}
The power cabinets of each building consist of two components, $(i)$ a smart switch controlled via ModBus/TCP, and $(ii)$ a virtual PLC handling the activation of the overcurrent protection.

The smart switch of a building measures the power load by summing constant and variable loads.
The former is the physiological energy consumption of the building, e.g., for running the video surveillance system.
The latter is the load that varies with some environmental conditions.
In particular, we consider occupants' equipment to be the main source of variable load.
Device are implemented through docker containers, and each container is labeled with its nominal energy consumption value, i.e., a metadata property called \texttt{load}.
Thus, variable load calculation amounts to the sum of the nominal energy consumption of all the containers associated with a certain building.
The association between a building and a container is also specified as a metadata property, namely \texttt{cabinet}.
The virtual switch exposes ModBus/TCP (read-only) registers for the current and maximum power consumption and coils for the master and trip commands. 
Also, the virtual switch described above is registered as ModBus/TCP slave of an OpenPLC instance.
In particular, its registers are mapped to input (I) addresses 100 and 101, while the trip switch coil is mapped to output (Q) address 100.\footnote{The master switch is not handled by the PLC, but is directly under the control of the SCADA system.}  
The PLC runs the following structured text program, which implements the current protection mechanism.
\begin{lstlisting}[basicstyle=\ttfamily\footnotesize,mathescape,morekeywords={let}]
PROGRAM trip               IF MAXCONS > 0 THEN
 VAR                        TRIPSW:= CONS > MAXCONS;
 CONS AT %IW100: WORD;     ELSE
 MAXCONS AT %IW101: WORD;   TRIPSW:= 0;
 TRIPSW AT %QX100: BOOL;   END_IF
END_VAR                   END_PROGRAM

\end{lstlisting}

The code given above compares the current power consumption (\texttt{MAXCONS}) with the rated one (\texttt{CONS}), only if the reported max consumption is a positive value. 
If this happens, the PLC sets the trip bit coil (\texttt{TRIPSW}) of the virtual switch to turn it off.
Such power trip is implemented by instructing the Docker API to stop the attached client containers.
\end{example}

For what concerns field device integration, we rely on Ditto.
Ditto acts as a centralized data aggregator for the devices it manages.
It provides a message broker that works by $(i)$ holding the latest reported state of each device and $(ii)$ forwarding messages, e.g., about issued commands. 
In Example~\ref{ex:ditto}, we describe how Ditto can emulate a field controller starting with the solar panel subsystem.

\begin{example}
\label{ex:ditto}
Consider again the solar panel simulation described in Example~\ref{ex:interpolation}.
We implement its field controller by means of the following Ditto HTTP endpoint.
\begin{center}
\begin{footnotesize}
\texttt{api/2/things/.../features/panel/properties/power}
\end{footnotesize}
\end{center}
The solar panel simulator detailed in Section~\ref{sec:physical} submits (method PUT) the current sensor values to Ditto every 10 seconds.
Then, the latest value is periodically retrieved by the SCADA system that queries (method GET) the Ditto HTTP endpoint given above.
\end{example}

\subsubsection{Network layer}
\label{sec:demo-network}
We use the predefined network chassis layout (see Section~\ref{sec:impl-network}) to implement the ICT network of the SPM (see Figure~\ref{fig:network}).
Briefly, chassis zones Industrial, Industrial/enterprise, Enterprise, and DMZ implement the field, control, client, and DMZ networks, respectively. 
A connection to the Internet is created by means of a virtual router playing the role of the ISP infrastructure. 
The ISP router is then attached to a host-through network which directly accesses the underlying host network and, thus, the Internet.

\subsubsection{Business logic layer}
\label{sec:demo-business}

\begin{figure*}
    \centering
    \includegraphics[width=\textwidth]{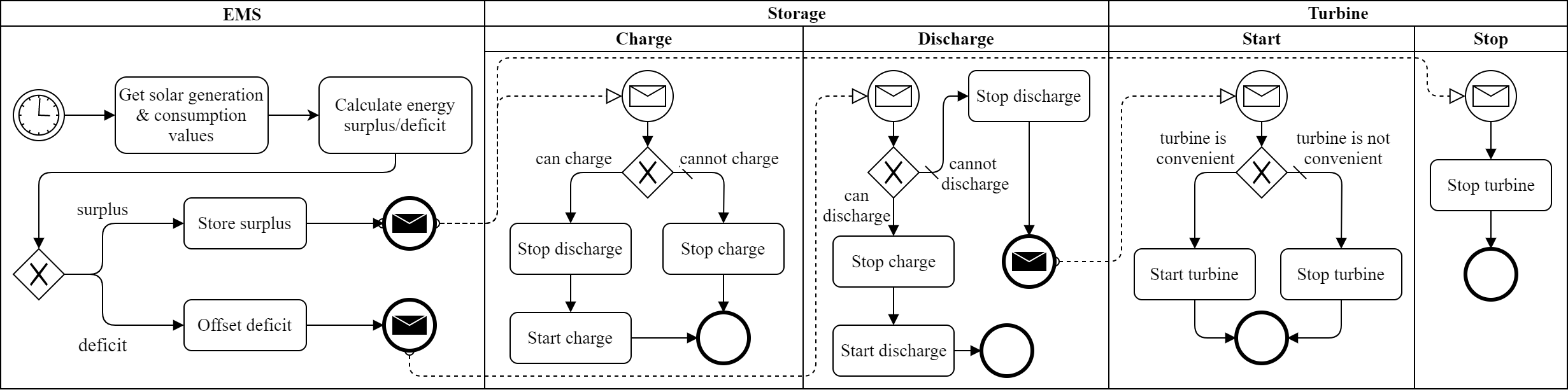}
    \caption{Business process of the Energy Management System}
    \label{fig:control}
\end{figure*}

The main process contributing to the business logic of the SPM is the EMS (see Section~\ref{sec:ems}).
We re-implement the EMS of~\cite{delfino-ems} with setpoint 0, i.e., our EMS aims at minimizing the amount of energy demanded to the utility network. The BPMN process for the EMS is given in Figure~\ref{fig:control}.
The overall process consists of five sub-processes, i.e., \emph{EMS}, \emph{Storage Charge}, \emph{Storage Discharge}, \emph{Turbine Start} and \emph{Turbine Stop}.
At regular intervals, EMS is activated by a timer event.
Initially, EMS queries the SCADA system for current solar energy generation and campus consumption.
These values are then compared to establish whether there is either a deficit or a surplus of energy.
In case of surplus, an event triggers both the Storage Charge and the Turbine Stop processes.
Instead, energy deficit results in an event activating the Storage Discharge process.
The Storage Charge process initially checks charging viability, i.e., whether the current level of charge is below a certain efficiency threshold, currently set to 90\%.
If this is the case, the process stops the discharge mode and starts charging the storage.
This enforces mutual exclusion between charge and discharge modes.\footnote{The actual storage system also supports them in parallel.}
When charging is not possible, the charge mode is disabled.
Symmetrically, Storage Discharge checks whether discharge mode is possible (with efficiency threshold set to 10\%). 
If so, the charge is stopped and discharge starts.
Otherwise, discharge is stopped and an event is fired for triggering Turbine Start. 
Turbine Start checks the activation conditions according to the system state and decides if switching the gas turbine on is \emph{convenient}, i.e., if the current deficit is more than the turbine efficiency threshold of 65 kW.
Depending on this check, either the start or stop command is issued.
Finally, the Turbine Stop process simply stops the turbine.

Each task appearing in the processes discussed above is implemented in Groovy.
Below we show the implementation of \emph{Get solar generation \& consumption values}.

\begin{lstlisting}[language=Java,basicstyle=\ttfamily\footnotesize,mathescape=false,keywordstyle=\color{blue}\bfseries,identifierstyle=\color{black},sensitive=false, stringstyle=\color{red}\ttfamily, showstringspaces=false, morekeywords={def}]
def http = Connectors.getConnector(HttpConnector.ID)
def response = http.createRequest().get()
  .url(scadaAPIURL + '/datapoint/getAll').execute()
def object = new JsonSlurper().parseText(response)
def points = [:]
object.each {
  def (pointName, pointXid) = [ it.name, it.xid ]
  points[pointName] = pointXid }
return points
\end{lstlisting}
Briefly, the code above retrieves all the datapoints by means of a SCADA API and stores them in the \textit{response} variable.
Such a variable is then parsed from the JSON format and converted to a key-value map, called \texttt{points}, which is eventually returned for the following task.

\subsubsection{Application layer}
\label{sec:demo-application}

\begin{figure}
    \centering
    \includegraphics[width=\columnwidth]{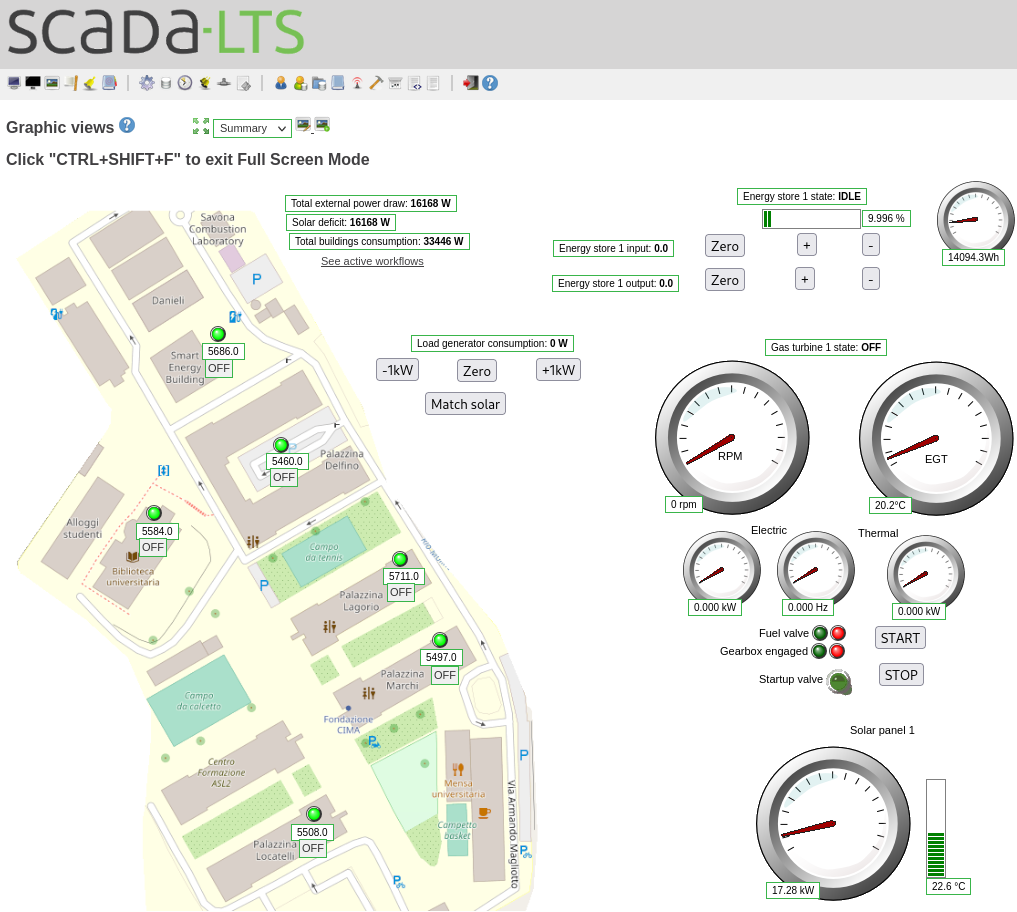}
    \caption{SPM operator interface implemented with Scada-LTS.}
    \label{fig:scadalts}
\end{figure}

One of the core applications hosted on the DT is the SPM SCADA system.
We use Scada-LTS~\cite{scada2021} to implement it.
Briefly, Scada-LTS consists of a Java servlet application running on top of an Apache Tomcat server.
Moreover, Scada-LTS interacts with an external MariaDB database for storing and retrieving collected data and general system information.
Scada-LTS resides in a docker container which is built out of the project source code via a Dockerfile.
Instead, an official, containerized version of MariaDB is directly downloaded from the Docker Hub repository.
The HMI of Scada-LTS is depicted in Figure~\ref{fig:scadalts}.

\subsubsection{Human layer}
\label{sec:demo-human}

At this layer we model the SPM turnout in terms of the behavior of students and staff members.
The total number $T$ of people accessing the university campus is available online (see~\cite{unige20conto} for staff, and~\cite{miur18conto} for students).
Following \cite[\S 3]{Zhan21building}, we assume that each person has two main effects, i.e., connecting a client to the campus network and, consequently, increasing the power consumption of a building.
We model the turnout by means of a template container which clusters $C$ individuals. 
Briefly, small values of $C$ increase the simulation granularity, while larger values improves the scalability. 
Thus, the buildings power consumption $E$ is computed as
$$
E = B + \sum_{i=0}^{T/C} C \cdot P_i
$$

\noindent
where $B$ is the base building consumption, i.e., the constant electrical load of the building equipment (e.g., safety lights), and $P_i$ is the individual power consumption following a normal distribution with mean $\mu_i$ and standard deviation $\sigma_i$, in symbols $P_i \sim N(\mu_i,\sigma_i)$.
The actual values used in our simulations are $C = 10$, $B = 1.5kW$, and $\forall i . \mu_i = 25$ and $\sigma_i = 5$.

\subsection{Execution and evaluation}
\label{sec:demo-evaluation}
Our execution environment runs on a Debian GNU/Linux, version 11, installed on a virtual machine hosted by VMWare ESXi 7.0U2 and configured with 16 Intel Xeon Gold 6252N at 2.3GHz, 64 GB of RAM, and 250GB of storage.
The running scenario includes 40 Docker containers implementing the DT elements discussed above.
Furthermore, depending on the occupancy of the building over time, 0 to 120 containers, i.e., client machines, spawn and connect to the enterprise network.
Finally, the workstation connected to the industrial/enterprise network runs a Microsoft Windows 10 OS on top of a containerized hypervisor (see Section~\ref{sec:applayer}) configured with 4 CPUs, 8 GB of RAM, and 60 GB of (dynamically allocated) disk. 

We used our DT for simulating one week of SPM activity.
During the execution, we collected time series of allocated system resources, e.g., CPU and memory, as well as SCADA data points.
The entire dataset is publicly available~\cite{dataset} and consists of 0.2B samples from the resource monitor and 8.4M SCADA samples.
Below we discuss some relevant facts for evaluating our simulation.

\begin{figure}
  \begin{center}
    \begin{tikzpicture}
     \begin{axis}[name=cpu,height=3cm, width=\columnwidth, xmin=0, xmax=168,
      grid=major, xticklabels=\empty,
      xtick={0, 24, 48, 72, 96, 120, 144, 168},
      yticklabel style = {font=\tiny},
      y label style={at={(0.08,0.52)}, font=\footnotesize},
      ylabel={RAM (GB)},
      ylabel={CPU (\%)}]
      \addplot[red]
        table[mark=none, col sep=comma] {csv/perf/cpu.csv};
      % avg
      \addplot[blue]coordinates {
        (0, 9.979039617295387)
    	(168, 9.979039617295387)
	    };
     \end{axis}

     \begin{axis}[name=ram, at={($(cpu.east)+(-7.31cm,-1.43cm)$)}, anchor=west, height=3cm, width=\columnwidth, xmin=0, xmax=168,
      grid=major, xticklabels=\empty,
      yticklabel style = {font=\tiny},
      y label style={at={(0.08,0.52)}, font=\footnotesize},
      ylabel={RAM (GB)},
      xtick={0, 24, 48, 72, 96, 120, 144, 168},
      ytick={ 18, 20, 22 }
      ]
      \addplot[red]
        table[mark=none, col sep=comma] {csv/perf/ram.csv}; 
      % avg
      \addplot[blue]coordinates {
        (0, 21.092182674105207)
    	(168, 21.092182674105207)
	    };
     \end{axis}

    %  \begin{axis}[name=net, at={($(ram.east)+(-7.28cm,-1.43cm)$)}, anchor=west, height=3cm, width=\columnwidth, xmin=0, xmax=168,
    %   grid=major, xticklabels=\empty]
    %   \addplot[red]
    %     table[mark=none, col sep=comma] {csv/perf/net_rx.csv}; 
    %   \addplot[blue]
    %     table[mark=none, col sep=comma] {csv/perf/net_tx.csv}; 
    %  \end{axis}

     \begin{axis}[name=disk, at={($(ram.east)+(-7.31cm,-1.43cm)$)}, anchor=west, height=3cm, width=\columnwidth, xmin=0, xmax=168,
      grid=major,
      ytick={0, 20, 40},
      xtick={0, 24, 48, 72, 96, 120, 144, 168},
      xticklabel style = {xshift=-0.5cm},
      yticklabel style = {font=\tiny},
      y label style={at={(0.08,0.47)}, font=\footnotesize},
      ylabel={Disk (GB)},
      xticklabels={{}, {Day1}, {Day2},
       {Day3}, {Day4}, {Day5}, {Day6}, {Day7}}
      ]
      \addplot[red]
        table[mark=none, col sep=comma] {csv/perf/disk_usage.csv}; 
     \end{axis}
    \end{tikzpicture}
    \caption{Computational load for the experiment.}
  \label{fig:cpumemedisk}
  \end{center}
\end{figure}
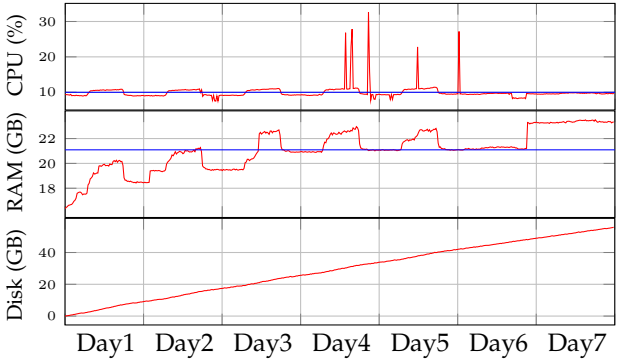

Figure~\ref{fig:cpumemedisk} shows the CPU, memory, and disk usage during the 7 days of execution.
The average CPU load was 10\% of the overall CPU capacity and never exceeded 31\%.
The simulation memory usage never exceeded 24GB.
The disk usage increased almost constantly by 345MB/h and the entire experiment generated 58 GB of data.
% The network traffic never exceeded 80 KB/s, considering both the load of DT networks and the host loopback.

Information about the computational resources needed for the simulation permit to estimate the affordability of our DT.
In particular, we considered the cost of hosting a Linux VM with sufficient computational resources in mainstream cloud providers.
Then, we evaluated the considered VM profile as 4 CPUs, 32GB RAM, and 100GB disk.
The prices\footnote{Last checked on \today} are $(i)$ 47.61 \$ on Google Cloud Platform, $(ii)$ 46.1 \$ on Amazon Web Services, and $(iii)$ 57,93 \$ on Microsoft Azure.
%42\$ GCP (24*7*0.25)
%46.75\$ AWS (187 monthly)
%74.47\$ Azure (297.89 monthly) 

%By leveraging SCADA samples, in Figure~\ref{fig:powerbal} we graph 
%\ernote{TODO: eval su dati SCADA}

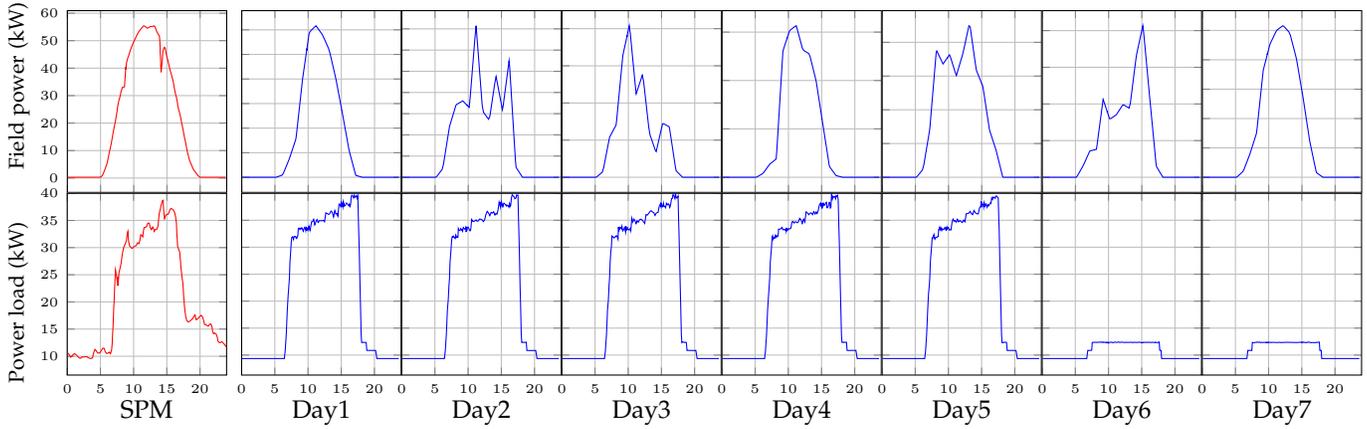
\begin{figure*}
  \begin{center}
     \begin{tikzpicture}
      \begin{axis}[ name=11, tiny,
        height=4cm, width=3.7cm, grid=major,
        xmin=0, xmax=24, id=major, xticklabels=\empty,
        y label style={at={(axis description cs:0.4, 0)},anchor=south},
        ylabel={\begin{tabular}{c @{\hspace{20pt}} r}{\footnotesize Power load (kW)} &{\footnotesize Field power (kW)}\\ \end{tabular}}]
        \addplot[red]
         table[mark=none, col sep=comma] {csv/SV/ppv.csv}; 
      \end{axis}

      \begin{axis}[ name=12, at={($(11.east)+(0.2cm,0)$)},anchor=west, tiny,     height=4cm, width=3.7cm, grid=major,
        xmin=0, xmax=24, id=major, xticklabels=\empty, yticklabels=\empty]
        \addplot
         table[mark=none, col sep=comma] {csv/data/DTpanel_12.csv}; 
      \end{axis}

      \begin{axis}[ name=13, at={($(12.east)+(0.01cm,0)$)},anchor=west, tiny,     height=4cm, width=3.7cm, grid=major,
        xmin=0, xmax=24, id=major, xticklabels=\empty, yticklabels=\empty]
        \addplot
         table[mark=none, col sep=comma] {csv/data/DTpanel_13.csv}; 
      \end{axis}

      \begin{axis}[ name=14, at={($(13.east)+(0.01cm,0)$)},anchor=west, tiny,     height=4cm, width=3.7cm, grid=major,
        xmin=0, xmax=24, id=major, xticklabels=\empty, yticklabels=\empty]
        \addplot
         table[mark=none, col sep=comma] {csv/data/DTpanel_14.csv}; 
      \end{axis}

      \begin{axis}[ name=15, at={($(14.east)+(0.01cm,0)$)},anchor=west, tiny,     height=4cm, width=3.7cm, grid=major,
        xmin=0, xmax=24, id=major, xticklabels=\empty, yticklabels=\empty]
        \addplot
         table[mark=none, col sep=comma] {csv/data/DTpanel_15.csv}; 
      \end{axis}
      
      \begin{axis}[ name=16, at={($(15.east)+(0.01cm,0)$)},anchor=west, tiny,     height=4cm, width=3.7cm, grid=major,
        xmin=0, xmax=24, id=major, xticklabels=\empty, yticklabels=\empty]
        \addplot
         table[mark=none, col sep=comma] {csv/data/DTpanel_16.csv}; 
      \end{axis}
      
      \begin{axis}[ name=17, at={($(16.east)+(0.01cm,0)$)},anchor=west, tiny,     height=4cm, width=3.7cm, grid=major,
        xmin=0, xmax=24, id=major, xticklabels=\empty, yticklabels=\empty]
        \addplot
         table[mark=none, col sep=comma] {csv/data/DTpanel_17.csv}; 
      \end{axis}
      
      \begin{axis}[ name=18, at={($(17.east)+(0.01cm,0)$)},anchor=west, tiny,      height=4cm, width=3.7cm, grid=major,
        xmin=0, xmax=24, id=major, xticklabels=\empty, yticklabels=\empty]
        \addplot
        table[mark=none, col sep=comma] {csv/data/DTpanel_18.csv}; 
      \end{axis}
      
      %% DT LOAD

      \begin{axis}[ name=11l, at={($(11.east)+(-2.12cm,-2.43cm)$)},anchor=west, tiny, height=4cm, width=3.7cm, grid=major,
        xmin=0, xmax=24, ymax=40, id=major, 
        x label style={at={(axis description cs:0.5, 0.13)},anchor=north},
        xlabel={\normalsize SPM}]
        \addplot[red]
        table[mark=none, col sep=comma] {csv/SV/pload.csv}; 
      \end{axis}

      \begin{axis}[ name=12l, at={($(11l.east)+(0.2cm,0)$)},anchor=west, tiny,      height=4cm, width=3.7cm, grid=major,
        xmin=0, xmax=24, ymax=40, id=major, yticklabels=\empty,
        x label style={at={(axis description cs:0.5, 0.13)},anchor=north},
        xlabel={\normalsize Day1} ]
        \addplot
        table[mark=none, col sep=comma] {csv/data/DTload_12.csv}; 
      \end{axis}

      \begin{axis}[ name=13l, at={($(12l.east)+(0.01cm,0)$)},anchor=west, tiny,      height=4cm, width=3.7cm, grid=major,
        xmin=0, xmax=24, ymax=40, id=major, yticklabels=\empty,
        x label style={at={(axis description cs:0.5, 0.13)},anchor=north},
        xlabel={\normalsize Day2} ]
        \addplot
        table[mark=none, col sep=comma] {csv/data/DTload_13.csv}; 
      \end{axis}

      \begin{axis}[ name=14l, at={($(13l.east)+(0.01cm,0)$)},anchor=west, tiny,      height=4cm, width=3.7cm, grid=major,
        xmin=0, xmax=24, ymax=40, id=major, yticklabels=\empty,
        x label style={at={(axis description cs:0.5, 0.13)},anchor=north},
        xlabel={\normalsize Day3}]
        \addplot
        table[mark=none, col sep=comma] {csv/data/DTload_14.csv}; 
      \end{axis}
      
      \begin{axis}[ name=15l, at={($(14l.east)+(0.01cm,0)$)},anchor=west, tiny,      height=4cm, width=3.7cm, grid=major,
        xmin=0, xmax=24, ymax=40, id=major, yticklabels=\empty,
        x label style={at={(axis description cs:0.5, 0.13)},anchor=north},
        xlabel={\normalsize Day4}]
        \addplot
        table[mark=none, col sep=comma] {csv/data/DTload_15.csv}; 
      \end{axis}      

      \begin{axis}[ name=16l, at={($(15l.east)+(0.01cm,0)$)},anchor=west, tiny,      height=4cm, width=3.7cm, grid=major,
        xmin=0, xmax=24, ymax=40, id=major, yticklabels=\empty,
        x label style={at={(axis description cs:0.5, 0.13)},anchor=north},
        xlabel={\normalsize Day5}]
        \addplot
        table[mark=none, col sep=comma] {csv/data/DTload_16.csv}; 
      \end{axis}

      \begin{axis}[ name=17l, at={($(16l.east)+(0.01cm,0)$)},anchor=west, tiny,      height=4cm, width=3.7cm, grid=major,
        xmin=0, xmax=24, ymax=40, id=major, yticklabels=\empty,
        x label style={at={(axis description cs:0.5, 0.13)},anchor=north},
        xlabel={\normalsize Day6}]
        \addplot
        table[mark=none, col sep=comma] {csv/data/DTload_17.csv}; 
      \end{axis}

      \begin{axis}[ name=18l, at={($(17l.east)+(0.01cm,0)$)},anchor=west, tiny,      height=4cm, width=3.7cm, grid=major,
        xmin=0, xmax=24, ymax=40, id=major, yticklabels=\empty,
        x label style={at={(axis description cs:0.5, 0.13)},anchor=north},
        xlabel={\normalsize Day7}]
        \addplot
        table[mark=none, col sep=comma] {csv/data/DTload_18.csv}; 
      \end{axis}
                  
    \end{tikzpicture}
    \caption{Comparison between DT with the photovoltaic field power and with the power absorbed by the load.}
  \label{fig:simulatedpower}
  \end{center}
\end{figure*}

% \begin{figure*}
%  \begin{center}
%   \begin{tikzpicture}
%     \begin{axis}[name=cpu, anchor=west, tiny, height=4cm, width=3.7cm, grid=major]
%     \addplot table[mark=none col sep=comma] {csv/perf/cpu.csv};
%     \end{axis}
%   \end{tikzpicture}
%  \end{center}
% \end{figure*}

Figure~\ref{fig:simulatedpower} shows the generated field power (above) and the total power load (below) during the DT simulation in comparison with one day of the same measures taken from the SPM.
Over the 7 days of simulation, the weather conditions influenced the actual generation of the solar panels. 
For instance, Day 6 and Day 7 show the performance during a cloudy day and a sunny one, respectively.
Although a direct comparison is not possible, the SPM record shows good correspondence with the profile of a sunny day.
In terms of power load, values shows the performance during both working days and the weekend (Day 6 and Day 7).
Again, there is a remarkable similarity between our simulation the actual load of the SPM during a working day.
Although more systematic measurements are needed to assess the accuracy of our DT, these comparisons already confirm the adequacy of our approach.
We plan to carry out further experiments as future work.

\paragraph*{\textbf{Discussion}} In terms of the requirements introduced in Section~\ref{sec:introduction}, we put forward the following statements.

\noindent
\textbf{General Purpose.} \toolname{} allows building heterogeneous and complex scenarios as in the SPM case study.
\smallskip

\noindent
\textbf{Expressive.} \toolname{} allows to implement assets, functionalities, and interactions related to all the six layers of the reference model. 
\smallskip

\noindent
\textbf{Extensible.} \toolname{} integrates several \emph{de facto} standards. 
Furthermore, it is the only DTF supporting the three technological pillars (i.e., simulation, emulation and passthrough). 
\smallskip

\noindent
\textbf{Affordable.} Our DTF prioritize affordable technologies and permits designers to rationally allocate the computational resources needed for the simulation.
\smallskip    

\noindent
\textbf{Lightweight.} By privileging lightweight emulation and simulation methods, \toolname{} significantly reduces the computational costs of simulations.

For all the reasons discussed above, we believe that LiDiTe can provide an open-source platform for fostering the development of DTs in many different contexts.

%% file: related.tex
\section{Related Work}
\label{sec:related}

Many authors have put forward definitions of DT in the last years (see~\cite{Tao2019} for a survey). 
Most of them focus on some specific aspects or technologies.
Although these proposals deal with some crucial aspects of real infrastructures, they do not introduce a general definition of DTF as we do in this paper.
Thus, here we focus on previous works proposing a general definition of DTF.
Table~\ref{tab:comparison} compares \toolname{} with the other DTF proposals.
There we report the supported levels and, for each of them, we indicate the enabling technologies among emulation, simulation, and passthrough.

The Electric Power and Intelligent Control twin (EPICTWIN)~\cite{Kandasamy21EPICTWIN} is a DTF for staging security training and research activities in a replica of a smart grid.
Interestingly, the DT model of EPICTWIN includes all the layers identified in this paper.
At the core of EPICTWIN stands Node-RED~\cite{openjs21node}, i.e., a visual programming tool wiring together hardware devices and software services. 
Instances of Node-RED are executed on Linux virtual machine to emulate the behavior of the plant at multiple layers.
For instance, Node-RED modules exist for running both the Field devices and the SCADA system, e.g., including an HMI application and the centralized control logic.
Like \toolname{}{}, network layer emulation is obtained by means of SDN. Instead, the Physical layer relies on passthrough for connecting external devices or third-party simulation software, e.g., Mathworks Simulink~\cite{simulink21}. 
Finally, EPICTWIN includes a module for the automatic execution of attack scripts.
Although partially, such a module simulates the activity of (hostile) human beings.

Eckhart et al. present CPS Twinning~\cite{Eckhart18Security}, a DTF aiming at mirroring cyber-physical systems (CPS).
Their proposal is inspired by MiniCPS~\cite{Antonioli15MiniCPS}, i.e, a framework for real-time CPS simulation.
They support both the integration with actuators/sensors and their simulation through historical values. 
Field devices mainly consist of emulated PLCs, executing standard ladder logic programs, but also passthrough of actual PLCs is supported.
The network layer emulation is implemented by Mininet~\cite{Lantz10Mininet}, which also supports passthrough.
Finally, at the Application layer, CPS Twinning emulates an HMI  with a command-line interface used to issue commands to PLCs.

The DT architecture presented in~\cite{redelinghuys2019six} aims at enabling the exchange of data between a remote emulation or simulation layer and the physical twin.
Their six-layer DT model extends the 5C architecture~\cite{Lee155C}, which provides the guidelines for developing CPSs in manufacturing application scenarios.
In particular, layers one and two represent devices and controllers of the physical twin.
Layer three connects elements of the lower layers using a vendor-neutral communication interface, i.e., the Open Platform Communications Unified Architecture (OPC UA)~\cite{opc21}, for retrieving and storing their data in a local repository.
Layer four represents the gateway to the IoT devices, and layer five enables the passthrough to a cloud-based repository for storing information of such devices.
The sixth layer includes emulation and simulation tools to analyze data stored in the lower layers, create reports, and make decisions for improving the performances of the physical twin processes.

Briefly, the layers three and four map to our Field layer, the layer five to our Application layer, and the layer six to our Control and Application layers.
The proposed DT implementation relies on a custom program for the IoT gateway and proprietary software for the other functionalities.

An alternative six-layer architecture for DTs to support human decision-making and AI-driven autonomy is introduced in~\cite{mostafa2020effective}.
Proceeding from bottom to top, the lowest layer, namely the Physical, is responsible for collecting data from the physical twin.
These data are normalized and stored by the two upper layers, i.e., Ingestion and Persistence, respectively. 
The Inference layer provides all data analytic functionalities, from simple formula-based calculations to machine learning algorithms.
The Service level implements interfaces to access data produced by the layers below.
It works as a gateway for the applications of the highest level, namely the Consumption layer, responsible for the monitoring, reporting, and interactive analytics.
All the functionalities of the above layers map to our Field and Application layers.
Moreover, the authors provide an implementation of the DT model as a case study for fault prediction in a production plant and rely only on open-source software. 

In~\cite{Behrendt19Open}, the authors also propose a microservices-based approach for designing and implementing DTs of smart manufacturing systems.
Their DTF has a five-layer structure inspired to the Reference Architecture Model for Industry 4.0 (RAMI 4.0)~\cite{hankel2015reference} and the Industrial Internet Reference Architecture (IIRA)~\cite{IIC2019}.
Similarly to our Field layer, in their approach lower layers acquire and aggregate data from edge devices.
Also, the highest levels integrate services for data monitoring and analysis, which we support at the Application layer.
Finally, the authors discuss which open-source software can be used to develop their DTF and provide an example implementation.

The toolkit presented in~\cite{Kamath20Ind} implements the DT model using only open-source platforms. 
In particular, it leverages Eclipse Hono~\cite{eclipse21hono} for integrating physical IoT devices via passthrough and, similarly to our proposal (see~\ref{sec:impl-field}), Eclipse Ditto~\cite{eclipse21ditto} for enabling their emulation.
Moreover, the toolkit integrates specific applications for data storage, data analytics, and visualization.

The White Label Digital Twins (WLDT)~\cite{Picone21WLDT} is proposed as a modular DTF.
Like the previous toolkit, it can be used for creating virtual instances that mirror IoT devices lying in the physical counterpart.
In general, WLDT can be seen as an alternative to Ditto and it aims to ensure more efficiency and flexibility.

\setlength{\tabcolsep}{4pt}

\begin{table}[t]
    \centering
    \scalebox{0.88}{
    \begin{tabular}{@{}lcc ccc ccc@{}}
         \toprule
           & \cite{Kandasamy21EPICTWIN} &  \cite{Eckhart18Security} & \cite{redelinghuys2019six} & \cite{mostafa2020effective} & \cite{Behrendt19Open} & \cite{Kamath20Ind} & \cite{Picone21WLDT} & \textbf{\toolname{}} \\
         \midrule
         {\sc P} & \nsimsymbol{}\nemusymbol{}\pstsymbol{} & \simsymbol{}\nemusymbol{}\pstsymbol{} & \simsymbol{}\nemusymbol{}\npstsymbol{} & & & & & \simsymbol{}\nemusymbol{}\pstsymbol{} \\
         {\sc F} & \nsimsymbol{}\emusymbol{}\npstsymbol{} & \nsimsymbol{}\emusymbol{}\pstsymbol{} & \nsimsymbol{}\nemusymbol{}\pstsymbol{} & \nsimsymbol{}\nemusymbol{}\pstsymbol{} & \nsimsymbol{}\nemusymbol{}\pstsymbol{} & \nsimsymbol{}\nemusymbol{}\pstsymbol{} & \nsimsymbol{}\nemusymbol{}\pstsymbol{} & \nsimsymbol{}\emusymbol{}\pstsymbol{} \\
         {\sc N} & \nsimsymbol{}\emusymbol{}\npstsymbol{} & \nsimsymbol{}\emusymbol{}\npstsymbol{} & & & & & & \simsymbol{}\emusymbol{}\pstsymbol{} \\
         {\sc B} & \nsimsymbol{}\emusymbol{}\npstsymbol{} & & \nsimsymbol{}\emusymbol{}\npstsymbol{} & & & & & \nsimsymbol{}\emusymbol{}\pstsymbol{}  \\
         {\sc A} & \nsimsymbol{}\emusymbol{}\npstsymbol{} & \nsimsymbol{}\emusymbol{}\npstsymbol{} & \nsimsymbol{}\emusymbol{}\pstsymbol{} & \nsimsymbol{}\emusymbol{}\npstsymbol{} & \nsimsymbol{}\emusymbol{}\npstsymbol{} & \nsimsymbol{}\emusymbol{}\npstsymbol{} & & \nsimsymbol{}\emusymbol{}\pstsymbol{} \\
         {\sc H} & \simsymbol{}\nemusymbol{}\npstsymbol{} & & & & & & & \simsymbol{}\nemusymbol{}\pstsymbol{} \\
         \bottomrule
    \end{tabular}
    }
    \caption{Comparison between DTF technologies (simulation \simsymbol{}, emulation \emusymbol{}, and passthrough \pstsymbol{}).}
    \label{tab:comparison}
\end{table}

%% file: conclusion.tex
\section{Conclusion}
\label{sec:conclusion}

In this paper, we introduced a novel DTF and its implementation, called \toolname{}. 
The distinguishing features of our proposal are $(i)$ the generality, flexibility and extensibility of the DTF, which includes six abstraction layers, and $(ii)$ an implementation which is based on available technologies and tools, and that has a very limited demand for computational resources.
Future directions include the integration of new technologies as well as the application to problems of interest, e.g., in the field of Cyber Ranges.

\section{Acknowledgment}
\label{sec:ack}

This work was partially funded by the Horizon 2020 project "Strategic Programs for Advanced Research and Technology in Europe" (SPARTA). 

%% file: main.bbl
\begin{thebibliography}{10}
\providecommand{\url}[1]{#1}
\csname url@rmstyle\endcsname
\providecommand{\newblock}{\relax}
\providecommand{\bibinfo}[2]{#2}
\providecommand\BIBentrySTDinterwordspacing{\spaceskip=0pt\relax}
\providecommand\BIBentryALTinterwordstretchfactor{4}
\providecommand\BIBentryALTinterwordspacing{\spaceskip=\fontdimen2\font plus
\BIBentryALTinterwordstretchfactor\fontdimen3\font minus
  \fontdimen4\font\relax}
\providecommand\BIBforeignlanguage[2]{{%
\expandafter\ifx\csname l@#1\endcsname\relax
\typeout{** WARNING: IEEEtran.bst: No hyphenation pattern has been}%
\typeout{** loaded for the language `#1'. Using the pattern for}%
\typeout{** the default language instead.}%
\else
\language=\csname l@#1\endcsname
\fi
#2}}

\bibitem{Fuller2020}
\BIBentryALTinterwordspacing
A.~Fuller, Z.~Fan, C.~Day, and C.~Barlow, ``Digital twin: Enabling
  technologies, challenges and open research,'' \emph{{IEEE} Access}, vol.~8,
  pp. 108\,952--108\,971, 2020. [Online]. Available:
  \url{https://doi.org/10.1109/access.2020.2998358}
\BIBentrySTDinterwordspacing

\bibitem{Wright2020}
\BIBentryALTinterwordspacing
L.~Wright and S.~Davidson, ``How to tell the difference between a model and a
  digital twin,'' \emph{Advanced Modeling and Simulation in Engineering
  Sciences}, vol.~7, no.~1, Mar. 2020. [Online]. Available:
  \url{https://doi.org/10.1186/s40323-020-00147-4}
\BIBentrySTDinterwordspacing

\bibitem{spm}
\BIBentryALTinterwordspacing
S.~Bracco, F.~Delfino, F.~Pampararo, M.~Robba, and M.~Rossi, ``The university
  of genoa smart polygeneration microgrid test-bed facility: The overall
  system, the technologies and the research challenges,'' \emph{Renewable and
  Sustainable Energy Reviews}, vol.~18, pp. 442--459, 2013. [Online].
  Available:
  \url{https://www.sciencedirect.com/science/article/pii/S1364032112005515}
\BIBentrySTDinterwordspacing

\bibitem{williams1998purdue}
T.~Williams, ``The purdue enterprise reference architecture and methodology
  (pera),'' \emph{Handbook of life cycle engineering: concepts, models, and
  technologies}, vol. 289, 1998.

\bibitem{hankel2015reference}
\BIBentryALTinterwordspacing
M.~Hankel and B.~Rexroth, ``{The Reference Architectural Model Industrie 4.0
  (RAMI 4.0)},'' \emph{ZVEI}, vol.~2, no.~2, pp. 4--9, 2015. [Online].
  Available:
  \url{https://przemysl-40.pl/wp-content/uploads/2010-The-Reference-Architectural-Model-Industrie-40.pdf}
\BIBentrySTDinterwordspacing

\bibitem{simulink21}
S.~Documentation, ``Simulation and model-based design,''
  \url{https://www.mathworks.com/products/simulink.html}, 2021.

\bibitem{Sontag98control}
E.~D. Sontag, \emph{{Mathematical Control Theory: Deterministic
  Finite-Dimensional Systems}}.\hskip 1em plus 0.5em minus 0.4em\relax
  Springer, 1998, {ISBN: 978-1-4612-0577-7}.

\bibitem{eclipse21ditto}
{Eclipse Foundation}, ``Ditto,'' \url{https://www.eclipse.org/ditto/}, accessed
  on July 2021.

\bibitem{openplc}
T.~R. Alves, M.~Buratto, F.~M. De~Souza, and T.~V. Rodrigues, ``Openplc: An
  open source alternative to automation,'' in \emph{IEEE Global Humanitarian
  Technology Conference (GHTC 2014)}.\hskip 1em plus 0.5em minus 0.4em\relax
  IEEE, 2014, pp. 585--589.

\bibitem{Naik2017}
\BIBentryALTinterwordspacing
N.~Naik, ``Choice of effective messaging protocols for {IoT} systems: {MQTT},
  {CoAP}, {AMQP} and {HTTP},'' in \emph{2017 {IEEE} International Systems
  Engineering Symposium ({ISSE})}.\hskip 1em plus 0.5em minus 0.4em\relax
  {IEEE}, Oct. 2017. [Online]. Available:
  \url{https://doi.org/10.1109/syseng.2017.8088251}
\BIBentrySTDinterwordspacing

\bibitem{Ramanathan2014TheI6}
R.~Ramanathan, ``The iec 61131-3 programming languages features for industrial
  control systems,'' \emph{2014 World Automation Congress (WAC)}, pp. 598--603,
  2014.

\bibitem{swales1999open}
A.~Swales \emph{et~al.}, ``{Open modbus/tcp specification},'' \emph{Schneider
  Electric}, vol.~29, pp. 3--19, 1999.

\bibitem{curtis2005dnp3}
K.~Curtis, ``{DNP3 Primer, Revision A},''
  \url{https://www.dnp.org/Portals/0/AboutUs/DNP3\%20Primer\%20Rev\%20A.pdf},
  accessed on July 2021.

\bibitem{openvpn21}
{OpenVPN}, ``{OpenVPN},'' https://openvpn.net/, 2021.

\bibitem{rfc7426}
\BIBentryALTinterwordspacing
E.~Haleplidis, K.~Pentikousis, S.~Denazis, J.~H. Salim, D.~Meyer, and
  O.~Koufopavlou, ``{Software-Defined Networking (SDN): Layers and Architecture
  Terminology},'' RFC 7426, Jan. 2015. [Online]. Available:
  \url{https://rfc-editor.org/rfc/rfc7426.txt}
\BIBentrySTDinterwordspacing

\bibitem{Bailey16Faucet}
\BIBentryALTinterwordspacing
J.~Bailey and S.~Stuart, ``Faucet: Deploying {SDN} in the enterprise,''
  vol.~14, no.~5, pp. 54--68, Oct. 2016. [Online]. Available:
  \url{https://doi.org/10.1145/3012426.3015763}
\BIBentrySTDinterwordspacing

\bibitem{188960}
\BIBentryALTinterwordspacing
B.~Pfaff, J.~Pettit, T.~Koponen, E.~Jackson, A.~Zhou, J.~Rajahalme, J.~Gross,
  A.~Wang, J.~Stringer, P.~Shelar, K.~Amidon, and M.~Casado, ``{The Design and
  Implementation of Open vSwitch},'' in \emph{12th {USENIX} Symposium on
  Networked Systems Design and Implementation ({NSDI} 15)}.\hskip 1em plus
  0.5em minus 0.4em\relax Oakland, CA: {USENIX} Association, May 2015, pp.
  117--130. [Online]. Available:
  \url{https://www.usenix.org/conference/nsdi15/technical-sessions/presentation/pfaff}
\BIBentrySTDinterwordspacing

\bibitem{openwrt21}
{OpenWrt Project}, ``{OpenWrt},'' https://openwrt.org/, 2021.

\bibitem{opnsense21}
{OPNsense Project}, ``{OPNsense},'' https://opnsense.org/, 2021.

\bibitem{international2010iec}
{International Electrotechnical Commission and others}, ``{IEC 62443:
  Industrial Communication Networks—Network and System Security},'' \emph{IEC
  Central Office: Geneva, Switzerland}, 2010.

\bibitem{reese2008nginx}
W.~Reese, ``Nginx: the high-performance web server and reverse proxy,''
  \emph{Linux Journal}, vol. 2008, no. 173, p.~2, 2008.

\bibitem{kvm}
A.~Kivity, Y.~Kamay, D.~Laor, U.~Lublin, and A.~Liguori, ``kvm: the linux
  virtual machine monitor,'' in \emph{Proceedings of the Linux symposium},
  vol.~1, no.~8.\hskip 1em plus 0.5em minus 0.4em\relax Dttawa, Dntorio,
  Canada, 2007, pp. 225--230.

\bibitem{compose21}
{Docker, Inc}, ``{Docker Compose},'' https://docs.docker.com/compose/, 2021.

\bibitem{white2004introduction}
S.~A. White, ``Introduction to bpmn,'' \emph{Ibm Cooperation}, vol.~2, no.~0,
  p.~0, 2004.

\bibitem{camunda2021}
``{Camunda},'' \url{https://camunda.com/}, 2021.

\bibitem{Bracco17smart}
S.~Bracco, F.~Delfino, F.~Foiadelli, and M.~Longo, ``{Smart microgrid
  monitoring: Evaluation of key performance indicators for a PV plant connected
  to a LV microgrid},'' in \emph{2017 IEEE PES Innovative Smart Grid
  Technologies Conference Europe (ISGT-Europe)}, 2017, pp. 1--6.

\bibitem{bracco-gas-turbine}
\BIBentryALTinterwordspacing
S.~Bracco and F.~Delfino, ``A mathematical model for the dynamic simulation of
  low size cogeneration gas turbines within smart microgrids,'' \emph{Energy},
  vol. 119, pp. 710--723, Jan. 2017. [Online]. Available:
  \url{https://doi.org/10.1016/j.energy.2016.11.033}
\BIBentrySTDinterwordspacing

\bibitem{arpal-clima}
{ Arpal }, ``{Atlante climatico della Liguria},''
  \url{https://www.arpal.liguria.it/contenuti_statici//clima/atlante/Atlante_climatico_della_Liguria.pdf},
  accessed on August 2021.

\bibitem{pvgis21}
{The European Commission's science and knowledge service}, ``{Photovoltaic
  Geographical Information System (PVGIS)},''
  \url{https://ec.europa.eu/jrc/en/pvgis}, 2021, accessed on August 2021.

\bibitem{delfino-ems}
\BIBentryALTinterwordspacing
F.~Delfino, M.~Rossi, F.~Pampararo, and L.~Barillari, \emph{An Energy
  Management Platform for Smart Microgrids}.\hskip 1em plus 0.5em minus
  0.4em\relax Berlin, Heidelberg: Springer Berlin Heidelberg, 2016, pp.
  207--225. [Online]. Available:
  \url{https://doi.org/10.1007/978-3-662-49179-9_10}
\BIBentrySTDinterwordspacing

\bibitem{scada2021}
``{Scada-LTS},'' \url{http://scada-lts.org/}, 2021.

\bibitem{unige20conto}
{Università degli Studi di Genova}, ``{Conto annuale del personale},''
  \url{https://unige.it/trasparenza/dotazioneorganica/conto\_annuale\_personale.shtml},
  2020, source is in Italian. Accessed on August 2021.

\bibitem{miur18conto}
{Ministero dell'Istruzione dell'Università e della Ricerca}, ``{Popolazione
  studentesca dell'Università degli Studi di Genov},''
  \url{http://ustat.miur.it/dati/didattica/italia/atenei-statali/genova}, 2018,
  source is in Italian. Accessed on August 2021.

\bibitem{Zhan21building}
\BIBentryALTinterwordspacing
S.~Zhan and A.~Chong, ``Building occupancy and energy consumption: Case studies
  across building types,'' \emph{Energy and Built Environment}, vol.~2, no.~2,
  pp. 167--174, 2021. [Online]. Available:
  \url{https://www.sciencedirect.com/science/article/pii/S2666123320300829}
\BIBentrySTDinterwordspacing

\bibitem{dataset}
\BIBentryALTinterwordspacing
G.~Longo, E.~Russo, and G.~Costa, ``{LiDiTE - SPM use case experiment},'' 2022.
  [Online]. Available: \url{https://data.mendeley.com/datasets/x3v2yhjx7c/1}
\BIBentrySTDinterwordspacing

\bibitem{Tao2019}
\BIBentryALTinterwordspacing
F.~Tao, H.~Zhang, A.~Liu, and A.~Y.~C. Nee, ``Digital twin in industry:
  State-of-the-art,'' \emph{{IEEE} Transactions on Industrial Informatics},
  vol.~15, no.~4, pp. 2405--2415, Apr. 2019. [Online]. Available:
  \url{https://doi.org/10.1109/tii.2018.2873186}
\BIBentrySTDinterwordspacing

\bibitem{Kandasamy21EPICTWIN}
\BIBentryALTinterwordspacing
N.~K. Kandasamy, S.~Venugopalan, T.~K. Wong, and L.~J. Nicholas, ``{EPICTWIN:}
  an electric power digital twin for cyber security testing, research and
  education,'' \emph{CoRR}, vol. abs/2105.04260, 2021. [Online]. Available:
  \url{https://arxiv.org/abs/2105.04260}
\BIBentrySTDinterwordspacing

\bibitem{openjs21node}
{OpenJS Foundation}, ``Node-red,'' \url{https://nodered.org/}, accessed on July
  2021.

\bibitem{Eckhart18Security}
\BIBentryALTinterwordspacing
M.~Eckhart and A.~Ekelhart, ``Towards security-aware virtual environments for
  digital twins,'' ser. CPSS '18.\hskip 1em plus 0.5em minus 0.4em\relax New
  York, NY, USA: Association for Computing Machinery, 2018, p. 61–72.
  [Online]. Available: \url{https://doi.org/10.1145/3198458.3198464}
\BIBentrySTDinterwordspacing

\bibitem{Antonioli15MiniCPS}
\BIBentryALTinterwordspacing
D.~Antonioli and N.~O. Tippenhauer, ``Minicps: A toolkit for security research
  on cps networks,'' in \emph{Proceedings of the First ACM Workshop on
  Cyber-Physical Systems-Security and/or PrivaCy}, ser. CPS-SPC '15.\hskip 1em
  plus 0.5em minus 0.4em\relax New York, NY, USA: Association for Computing
  Machinery, 2015, p. 91–100. [Online]. Available:
  \url{https://doi.org/10.1145/2808705.2808715}
\BIBentrySTDinterwordspacing

\bibitem{Lantz10Mininet}
\BIBentryALTinterwordspacing
B.~Lantz, B.~Heller, and N.~McKeown, ``A network in a laptop: Rapid prototyping
  for software-defined networks,'' in \emph{Proceedings of the 9th ACM SIGCOMM
  Workshop on Hot Topics in Networks}, ser. Hotnets-IX.\hskip 1em plus 0.5em
  minus 0.4em\relax New York, NY, USA: Association for Computing Machinery,
  2010. [Online]. Available: \url{https://doi.org/10.1145/1868447.1868466}
\BIBentrySTDinterwordspacing

\bibitem{redelinghuys2019six}
A.~Redelinghuys, A.~H. Basson, and K.~Kruger, ``A six-layer architecture for
  the digital twin: a manufacturing case study implementation,'' \emph{Journal
  of Intelligent Manufacturing}, pp. 1--20, 2019.

\bibitem{Lee155C}
\BIBentryALTinterwordspacing
J.~Lee, B.~Bagheri, and H.-A. Kao, ``A cyber-physical systems architecture for
  industry 4.0-based manufacturing systems,'' \emph{Manufacturing Letters},
  vol.~3, pp. 18--23, 2015. [Online]. Available:
  \url{https://www.sciencedirect.com/science/article/pii/S221384631400025X}
\BIBentrySTDinterwordspacing

\bibitem{opc21}
{OPC Foundation}, ``Opc ua online reference,''
  \url{https://reference.opcfoundation.org/ }, accessed on July 2021.

\bibitem{mostafa2020effective}
F.~Mostafa, L.~Tao, and W.~Yu, ``An effective architecture of digital twin
  system to support human decision making and ai-driven autonomy,''
  \emph{Concurrency and Computation: Practice and Experience}, p. e6111, 2020.

\bibitem{Behrendt19Open}
V.~Damjanovic-Behrendt and W.~Behrendt, ``An open source approach to the design
  and implementation of digital twins for smart manufacturing,''
  \emph{International Journal of Computer Integrated Manufacturing}, vol.~32,
  no. 4-5, pp. 366--384, 2019.

\bibitem{IIC2019}
\BIBentryALTinterwordspacing
{Industrial Internet Consortium (IIC)}, ``{The Industrial Internet of Things
  Volume G1: Reference Architecture},'' 2019. [Online]. Available:
  \url{https://www.iiconsortium.org/pdf/IIRA-v1.9.pdf}
\BIBentrySTDinterwordspacing

\bibitem{Kamath20Ind}
V.~Kamath, J.~Morgan, and M.~I. Ali, ``Industrial iot and digital twins for a
  smart factory : An open source toolkit for application design and
  benchmarking,'' in \emph{2020 Global Internet of Things Summit (GIoTS)},
  2020, pp. 1--6.

\bibitem{eclipse21hono}
{Eclipse Foundation}, ``Hono,'' \url{https://www.eclipse.org/hono/}, accessed
  on July 2021.

\bibitem{Picone21WLDT}
\BIBentryALTinterwordspacing
M.~Picone, M.~Mamei, and F.~Zambonelli, ``Wldt: A general purpose library to
  build iot digital twins,'' \emph{SoftwareX}, vol.~13, p. 100661, 2021.
  [Online]. Available:
  \url{https://www.sciencedirect.com/science/article/pii/S2352711021000066}
\BIBentrySTDinterwordspacing

\end{thebibliography}
